\documentclass[a4paper,fleqn,usenatbib]{mnras}

\usepackage{newtxtext,newtxmath}
\usepackage[T1]{fontenc}
\usepackage{ae,aecompl}

\usepackage{longtable}
\usepackage{natbib}
\usepackage{psfrag}
\usepackage{xspace}
\usepackage{color}
\usepackage{graphicx}   % Including figure files
\usepackage{amsmath}    % Advanced maths commands
\newcommand{\Rpl}{$R_{\rm pl}$\,}
\newcommand{\Mpl}{$M_{\rm pl}$\,}
\newcommand{\Rer}{\ensuremath{R_{\oplus}}\,}
\newcommand{\Mer}{\ensuremath{M_{\oplus}}\,}

\newcommand{\Msun}{M$_{\odot}$\,}
\title[Host star vs planetary population]{How does the mass and activity history of the host star affect the population of low-mass planets?}

\author[D. Kubyshkina et al.]{
Daria Kubyshkina,$^{1}$\thanks{E-mail: kubyshkd@tcd.ie} Aline
A.~Vidotto,$^{1}$ %Luca Fossati$^{2}$ and Eoin Farrell$^{1}$
\\
$^{1}$School of Physics, Trinity College Dublin, the University of Dublin, College Green, Dublin-2, Ireland\\
}

\date{Accepted XXX. Received YYY; in original form ZZZ}

\pubyear{2021}

\begin{document}
\label{firstpage}
\pagerange{\pageref{firstpage}--\pageref{lastpage}} \maketitle

\begin{abstract}
%it is 202 words of 250 now
The evolution of the atmospheres of low and intermediate-mass planets is strongly connected to the physical properties of their host stars. The types and the past activities of planet-hosting stars can, therefore, affect the overall planetary population. In this paper, we perform a comparative study of sub-Neptune-like planets orbiting stars of different masses and different evolutionary histories. We discuss the general patterns of the evolved population as a function of parameters and environments of planets. As a model of the atmospheric evolution, we employ the own framework combining planetary evolution in MESA with the realistic prescription of the escape of hydrogen-dominated atmospheres. {We find that the final populations look qualitatively similar in terms of the atmospheres survival around different stars, but qualitatively different, with this difference accentuated for planets orbiting more massive stars. We show that a planet has larger chances of keeping its primordial atmosphere in the habitable zone of a solar mass star compared to M or K dwarfs and if it starts the evolution having a relatively compact envelope.} We also address the problem of the uncertain initial temperatures (luminosities) of planets and show that this issue is only of particular importance for planets exposed to extreme atmospheric mass-losses.
\end{abstract}

\begin{keywords}
Hydrodynamics -- Planets and satellites: atmospheres -- Planets
and satellites: physical evolution
\end{keywords}

%%%%%%%%%%%%%%%%%%%%%%%%%%%%%%%%%%%%%%%%%%%%%%%%%%

%%%%%%%%%%%%%%%%% BODY OF PAPER %%%%%%%%%%%%%%%%%%

\section{Introduction}\label{sec::intro}

The evolution of planetary hydrogen-dominated atmospheres is controlled by the thermal evolution of the planet \citep[e.g.,][]{rogers2010,nettelmann2011,miller2011,valencia2013,lopez2012,lopez2014}, its atmospheric mass loss \citep[e.g.,][]{watson1981,lammer2003,lammer2016,Erkaev2007,erkaev2016,Lecavelier2004,owen2016,salz2016,kubyshkina2018grid,gupta2019,gupta2020}, and is strongly dependent on the stellar environment around the planet. This includes the heating by the host star and the amount of high-energy (X-ray and extreme ultraviolet, XUV) radiation received by the planet \citep[e.g.,][]{kubyshkina2018grid}. Both depend on the type of the host star (in particular, its mass and temperature) and the orbital distance of the planet. In addition, the XUV luminosity of the star of a given mass changes through time not univocally. Instead, the X-ray (and hence, XUV) luminosities of stars with similar masses in young clusters show a large spread up to about an order of magnitude due to the difference in the initial rotation rates of stars \citep[e.g.,][]{wright2011,tu2015,magaudda2020,Johnstone2020mors}. This uncertainty in the stellar XUV holds up to about 1~Gyr, i.e., for the period when stars are most active and planets are hot and inflated after the dispersal of the protoplanetary disk, which is also when atmospheric escape is strongest \citep[e.g.,][]{owen2016,kubyshkina2018grid,gupta2019,gupta2020}. Thus simultaneous consideration of stellar and planetary evolution becomes important, and the population of evolved planets can look essentially different around stars of different types.

Another issue that may play a role in the atmospheric evolution is the initial state of a planet at the protoplanetary disk dispersal when atmospheric evaporation begins. Except for the parameters that can often be considered nearly unchanged after this moment (as the total mass of the planet and its orbit), other important initial parameters remain unknown, such as the initial atmospheric mass fraction and post-formation luminosity (hence, temperature) of a planet. Both of these parameters can be predicted by planetary formation and accretion models \citep[e.g.,][]{mizuno1980,mordasini2012,mordasini2017,leconte2015,morbidelli2009,morbidelli2016,morbidelli2020}. Their specific values for a given planet, however, depend on the formation scenario of a planet and on the properties of a protoplanetary disk, which are in general unknown, and on the assumptions of the specific model. This leads to large uncertainties in the prediction of some initial planetary parameters (see, e.g., \citealt{ikoma2012} for atmospheric masses and \citealt{mordasini2017} for initial luminosities).

In spite of this, certain initial parameters can be constrained by the present-day parameters of a planet. For the initial atmospheric mass fraction, this can be done by fitting the observed parameters of a planet, considering the evolution of the planetary atmosphere in a statistical way. This requires, however, both planetary radius and mass to be measured with sufficient accuracy, and the planet should not be exposed to extreme atmospheric escape throughout the evolution \citep[][]{kubyshkina2019b}. Considering initial luminosities (temperatures) of planets, \citet{owen2020} has shown that the post-formation luminosities can be constrained for young planets using escape argument, but only if the planets are not older than a few tens of Myr. We further address the issue of unknown post-formation luminosities here.

In the present paper, we focus on how the evolution of atmospheres of low-mass planets (with masses smaller than Neptune) is affected by the mass and activity history of the host stars. It was shown by \citet{Cloutier2020}, that the population of planets does not look the same around M dwarfs in comparison to heavier stars. In particular, they show that the relative occurrence rate of rocky to non-rocky planets increases $\sim$6-30 times around mid-M dwarfs compared to mid-K dwarfs, and the slope of the radius valley bears the opposite sign. This can suggest that the atmospheric photoevaporation (the part of atmospheric mass loss related to the stellar XUV irradiation) may be more effective around low-mass stars. Here, we simulate planets evolving around M (0.5~$M_{\odot}$), K (0.75~$M_{\odot}$) and G (1.0~$M_{\odot}$) dwarfs, accounting for the possible variations in their activity history. We find that atmospheric escape of planets with the same equilibrium temperature ranges occurs more efficiently around lower-mass stars despite their luminosities being smaller. This indirectly supports the findings of \citet{Cloutier2020}. Also, for low-mass stars, we find the maximum differences in atmospheric mass loss caused by various activity histories are smaller than those caused by different stellar masses.

For our model, we use the upgraded version of the framework developed in \citet{kubyshkina2020mesa}, which combines the thermal evolution of planetary atmospheres simulated by MESA (Modules for Experiments in Stellar Astrophysics, \citealt{paxton2018}) with the realistic prescription of the atmospheric mass loss from \citet{kubyshkina2018approx}. We now employ more sophisticated models to describe an input from the host star, i.e., the heating of planets and XUV irradiation. While in the previous version we considered a single model assuming solar-like host star with the XUV luminosity changing with time according to the power-law approximation by \citet{ribas2005}, in the present version, we employ the set of stellar models by \citet{Johnstone2020mors}. The latter allows to consistently model stellar XUV luminosity throughout the evolution for different stellar masses and allows to include various stellar activity histories. %

This paper is organized as follows. In Section \ref{sec::model} we describe the modeling approach employed in this paper and the stellar models we use. In Section~\ref{ssec:model-entropy}, we derive what is the possible range of initial (post formation) planetary luminosities and the possible effects this poorly known parameter has on the final properties of evolved planets. In Section~\ref{sec::results}, we present and discuss the results of the simulation for a set of planets evolving in different stellar environments. In Section~\ref{ssec::results-overview} we give an overview of the results, and in Section~\ref{ssec::results-tau} we discuss the survival timescales of the atmospheres of planets with specific initial parameters. In Section~\ref{sec::approximation} we propose an analytical approximation of the relation between the atmospheric mass fraction and planetary parameters. Finally, in Section~\ref{sec::discussion} we summarize our conclusions.

%%%%%%%%%%%%%%%%%%%%%%%%%%%%%%%%%%%%%%%%%%%%
\section{The modelling approach.}\label{sec::model}

\subsection{Thermal evolution and  atmospheric escape in MESA.}\label{sec::model_overview}

The basic approach employed in this work was described in detail
in \citet{kubyshkina2020mesa}. In brief, we use the `Modules for
Experiments in Stellar Astrophysics' (MESA)
\citep{paxton2011,paxton2013,paxton2018} to couple the thermal
evolution of a planet (lower atmosphere) with the realistic
prescription of the atmospheric escape. The former is computed
through the own planetary framework of MESA \citep{paxton2018},
and as the latter we use the hydro-based approximation (HBA)
\citep{kubyshkina2018approx} based on the grid of hydrodynamical
models of planetary upper atmospheres \citep{kubyshkina2018grid}.

In the present work, we test and update the initial conditions for
the newborn planets as described in
Section~\ref{ssec:model-entropy}. The grid of considered planetary
masses and initial atmospheric mass fractions remains the same as
in the previous work \citep{kubyshkina2020mesa}: we study planets
in mass range of 5-20 Earth mass (\Mer) and initial atmospheric
mass fractions of 0.5-30$\%$. Different from the previous work,
the range of planetary masses corresponds to the total masses of
planets instead of the core masses. {This change does not have
significant effects in the results and was introduced to unify the
masses of planets with different initial atmospheric mass
fractions at the beginning of the evolution, when the atmospheric
mass loss is strongest. In the same mass bin, we compare therefore
the planets with equivalent masses instead of those differing up
to one third of their mass (between 0.5 and 30\% of the envelope
around the same core), as it was in the previous version of the
model.}

These planets evolve at different orbits around stars with
different masses (0.5, 0.75, and 1.0 solar mass) and activity
histories as described in \ref{ssec::model-Mors}. We consider the
orbital separations between 0.03 and 0.5 AU, excluding the orbits
where equilibrium temperatures are below 300 K, which is the
restriction of our atmospheric escape models
\citep{kubyshkina2018grid,kubyshkina2018approx}.

The basic algorithm we employ to set up the model planet in MESA
consists of five general steps: (1) creation of a coreless planet
with mass of 0.1 Jupiter mass; (2) adding the solid core (i.e.,
setting the lower boundary conditions); (3) reducing the
atmosphere to the desired value; (4) standardizing the energy
budget through setting the initial entropy (i.e., adjusting the
lower boundary conditions to stay consistent within the considered
set of planets); (5) setting up the stellar heating (i.e., setting
the upper boundary conditions) and letting the planet evolve for
5~Myr without atmospheric escape.

{Except for step 4, all the other steps were described in depth in
\citet{kubyshkina2020mesa}. The initial entropy there was,
however, set more or less arbitrarily, as this parameter is poorly
constrained for young planets. For this reason, we explore this
issue in more details in Section~\ref{ssec:model-entropy}. With
minor proviso we confirm that the particular choice of the initial
entropy is of less importance for the final state of our simulated
planets, thus the readers who are more interested in the
post-evolution planetary distributions can skip
Section~\ref{ssec:model-entropy}.}

\subsection{Different stellar masses and evolutionary paths.}\label{ssec::model-Mors}

The rotation of the star and its high-energy radiation (X-ray plus
extreme ultraviolet, XUV) are closely connected, with faster
rotating stars being XUV brighter, and both rotation and XUV
luminosity decaying with time. This evolution though does not
follow a unique path and depends on the stellar parameters, in
particular stellar mass and the initial rotation rate; the
rotation rates at the young ages ($\sim$ earlier than 1\,Gyr)
present a wide spread, and at the later stage the evolutionary
tracks converge to the same path for a specific stellar mass
\citep[e.g.,][]{tu2015,Johnstone2020mors}.

In the present work we consider three stellar masses of 0.5, 0.75
and 1.0 solar mass ($M_{\odot}$). To model their rotation and XUV
evolution, we employ the models developed by
\citet{Johnstone2020mors}\footnote{The MORS (MOdel for Rotation of
Stars) code is available at https://ColinPhilipJohnstone/Mors.}.
This model represents the physical rotation model constrained by
observations in young stellar clusters. To describe the internal
stellar physics, it employs the evolutionary models by
\citet{Spada2013} and use observations to derive empirical
relations between stellar XUV luminosity and rotation for
different stellar masses between 0.1 and 1.25\,$M_{\odot}$.

To account for the whole range of possible rotation histories of
stars, for each of stellar masses we consider the `fast' and `slow
rotator' scenarios, assigning the rotation periods at 150~Myr to
0.5/1 and 7 days, respectively. We take for the fast rotator 0.5
day for stellar masses of 0.5 and 0.75\,\Msun and 1 day for
1.0\,\Msun where the latter is dictated by the model restrictions.
These periods correspond to the short and long period wings in the
distribution observed in young clusters (of $\sim$ 150~Myr old)
\citep[e.g.,][]{johnstone2015rot,Johnstone2020mors}. {When further
in the paper we address `fast' and `slow' rotating stars, we mean
these exact models.}

\begin{figure}
  \includegraphics[width=\hsize]{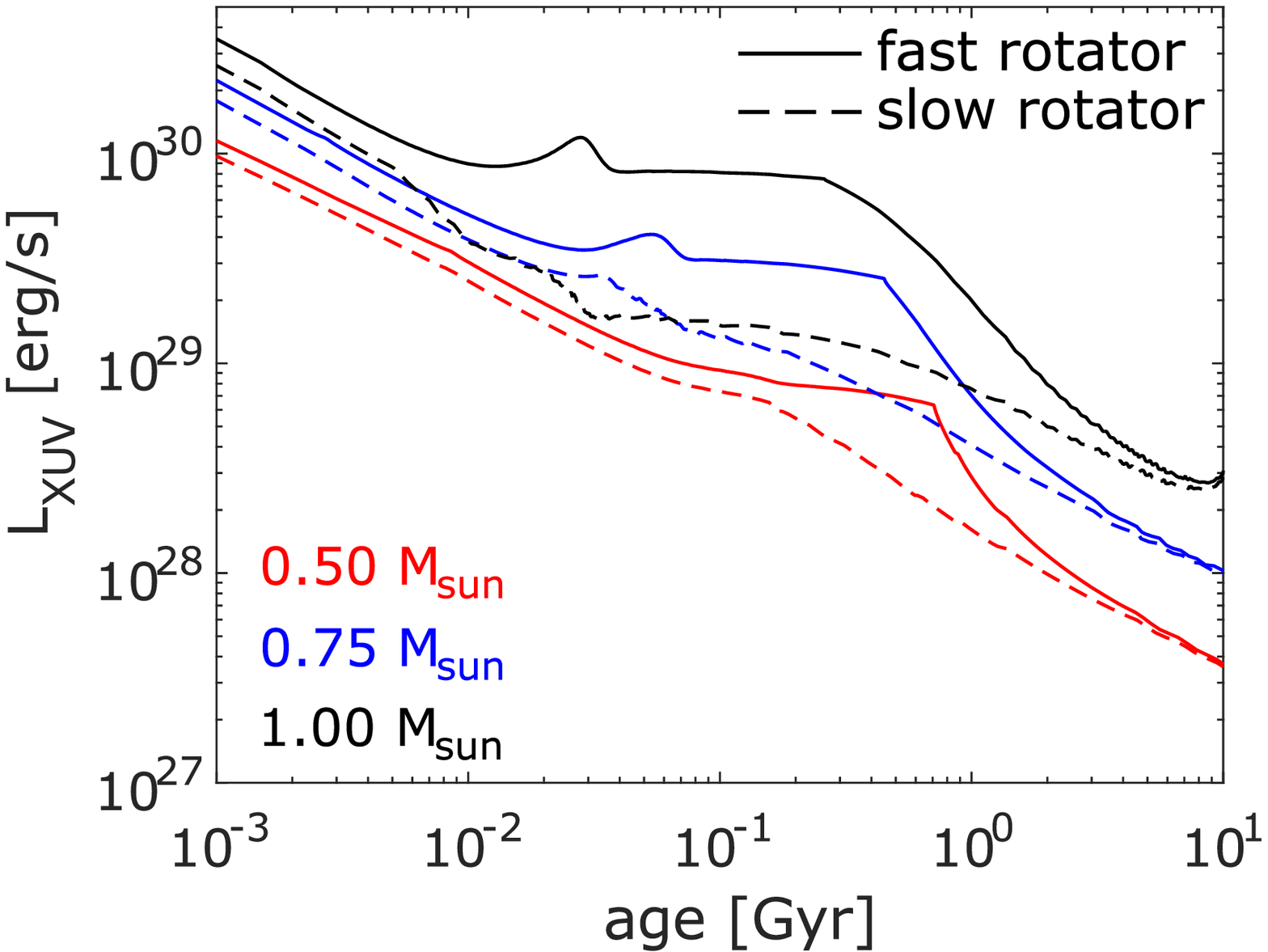}\\
  \includegraphics[width=\hsize]{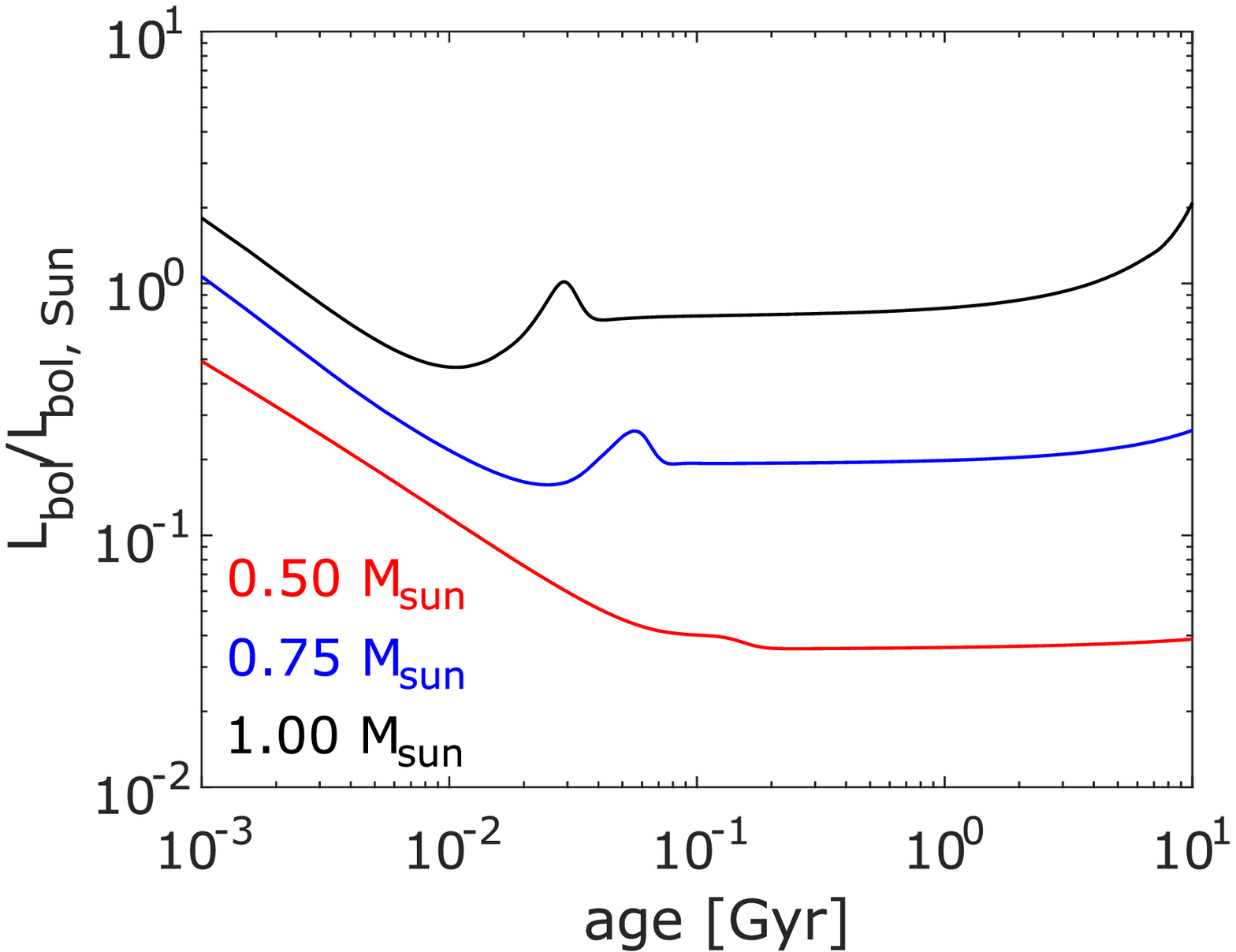}\\
  \caption{Evolution of the stellar XUV (top panel) and bolometric (bottom panel) luminosities as given by the Mors code for stars of 0.5 (red), 0.75 (blue) and 1.0 (black) solar mass, evolving as fast (solid lines) or slow (dashed lines) rotators.}\label{fig::mors_tracks}
\end{figure}

The resulting XUV luminosity ($L_{\rm XUV}$) evolution tracks and
corresponding bolometric luminosities ($L_{\rm bol}$) are
presented in Figure~\ref{fig::mors_tracks}. The largest XUV
radiation, as well as the largest difference between the slow and
the fast rotators, are seen for the solar mass star. For each
stellar mass, the XUV evolution for different rotators starts with
nearly the same $L_{\rm XUV}$ in the saturation regime (defined by
the saturation rotation period). The slow rotators drop out of the
saturation regime earlier (with average saturation time being
longer for low-mass stars), creating a bifurcation, and the tracks
converge again after $\sim$1~Gyr.

To stay consistent with the new XUV model, we change our
prescription of the stellar heating \citep[previously based
on][]{choi2016} to the equilibrium temperatures that are
calculated from the stellar evolution tracks from
\citet{Spada2013}.

\section{Initial conditions: post-formation luminosity of young planets}\label{ssec:model-entropy}

While other stellar and planetary parameters are measurable, the
initial (post-formation) luminosity of the planet, hence entropy,
is uncertain. Its value depends on the planetary mass and on a
range of processes throughout the formation stage, as pebble and
planetesimal accretion, gas accretion onto the core, contraction
of the envelope, decay of radioactive elements in the core, etc,
and also on the chosen formation scenario. These processes have
been studied extensively for hot Jupiters \citep[see,
e.g.,][]{bodenheimer2000,fortney2005,fortney2008,marley2007,spieg_burr2012,mordasini2012,mordasini2013,mordasini2017,owen_menou2016,berardo2017,marleau2017}.
Although these processes are relatively well known, depending on
the particular formation history, the final post-formation
luminosity of the core of a specific mass may vary by up to about
two orders of magnitude.

For sub-Neptune-like planets, this question was not addressed a
lot so far. \citet{owen2020} showed that, after the age of
$\sim$100\,Myr, one can not put a tight constraint on the initial
luminosity of the planet based on its present parameters. This
suggests that the effect of the uncertain post-formation
luminosity on planetary radius is short-living and therefore its
choice is not critical for the study of the evolved population of
planets. To look into this effect in more detail, we consider here
three planets: the very young AU Mic b ($\sim$ 22~Myr) and K2-33 b
($\sim$ 9~Myr), and slightly older Kepler-411 d ($\sim$ 212~Myr),
and see how their initial luminosities affect their evolution. The
main planetary and stellar parameters are summarized in
Table~\ref{tab::young_and_fit}. For K2-33 b, only the conservative
upper limit for the mass is available. The size of the planet,
however, suggests that the true mass is likely to be somewhat
smaller than Jupiter ($\leq 318 M_{\oplus}$). In the present work,
we follow the approach used by \citet{kubyshkina2018k2-33} and
consider in our simulations the upper mass limit of
40$M_{\oplus}$, which is the maximum mass allowing for usage of
HBA for the atmospheric mass loss. As we show later, this upper
limit is sufficient for the current task. We further exclude from
consideration masses smaller than 10\Mer; as we have shown in
\citet{kubyshkina2018k2-33} by direct hydrodynamic modeling, such
low mass would not allow K2-33 b to sustain any atmosphere at the
age of the system.

\begin{table*}
\caption{Key parameters of the young systems considered in
section~\ref{ssec:model-entropy}.}
\begin{tabular}{|c|c|c|c|}
  \hline
  parameter                & AU Mic b              & K2-33 b                          & Kepler-411 d \\
  \hline
  $M_*  [M_{\odot}]$           & $0.5\pm0.03^{(a)}$    & $0.56\pm 0.09^{(d)}$             & $0.870\pm0.039^{(f)}$ \\
  $R_*  [R_{\odot}]$           & $0.75\pm0.03^{(a)}$   & $1.05\pm 0.07^{(d)}$             & $0.82\pm0.02^{(f)}$ \\
  $P_{\rm rot,*}$ [day]        & $4.863\pm0.010^{(a)}$ &  $6.27\pm0.17^{(d)}$             & $\geq 9.78^{(g)}$ \\
  age   [Myr]                  & $22\pm3^{(a)}$        & $9.3_{-1.3}^{+1.1 (d)}$          & $212\pm31^{(f)}$ \\
  $L_{EUV}$ [$10^{28}$ erg/s]  & $5.85^{(b)}$          & $30.1^{(d,e)}$                   & $9.35^{(g,e)}$\\
  $d_0$ [AU]                   & $0.066\pm0.007^{(a)}$ & $0.0409_{-0.0023}^{+0.0021 (d)}$ & $0.279\pm0.004^{(f)}$ \\
  \Mpl  [\Mer]                 & $16.7\pm4.9^{(c)}$    & $\leq1176^{(d)}$                 & $15.2\pm5.1^{(f)}$ \\
  \Rpl  [\Rer]                 & $4.198\pm0.202^{(a)}$ & $5.04_{-0.37}^{+0.34 (d)}$       & $3.319\pm0.104^{(f)}$ \\
  \hline
\end{tabular}
\footnotesize{\\The sources of the data in Table are (a) --
\citet{plavchan2020}; (b) -- \citet{chadney2015}; (c) --
\citet{carolan2020}/Plavchan et al. (in prep.); (d) -- \citet{mann2016}; (e) --
\citet{Johnstone2020mors}; (f) -- \citet{sun2019}; (g) --
\citet{xu2020}.} \label{tab::young_and_fit}
\end{table*}

For a given planet with a known mass, the radius is defined by the
atmospheric mass fraction and the temperature profile of the
atmosphere, where the latter is controlled by the internal
luminosity of the planet and the stellar heating. For younger
planets, AU Mic b and K2-33 b, we fit the observed parameters
(i.e., their radii for the given planetary masses and positions
relative to the host stars) at a given age using the MESA
framework. To do so, we set various atmospheric mass fractions
keeping the mass of the planet constant, then relaxed stellar
irradiation to the level provided by observations and models by
\citet{Johnstone2020mors}, and finally adjust the central (core)
entropy of the planet (hence luminosity) to fit the observed
radius. In the following, we will in general refer to the entropy
of the planet rather than luminosity. This is due to the entropy
being a more general parameter and remaining in a similar range
for planets considered here, while the more intuitive planetary
luminosity varies significantly. Those two parameters are related
with each other and other planetary parameters through the
equation of state \citep{saumon1995}, and we show these relations
in Appendix~\ref{apx::S-L}.

In the present consideration, the lower mass limits for AU Mic b
and K2-33 b are nearly the same, and the upper mass limit for
K2-33 b is about twice the one for AU Mic b. The ages of the two
systems are similar. However, the EUV luminosity of K2-33 is about
5 times higher than the one of AU Mic \citep[which appears to have
relatively low activity in comparison to other stars of a similar
age, see, e.g., measurements in young clusters
in][]{Johnstone2020mors}. Due to this, and the slightly closer
orbital separation, K2-33 b experiences a much stronger
atmospheric escape for a very similar range of masses and
atmospheric mass fractions.

\begin{figure}
  \includegraphics[width=\hsize]{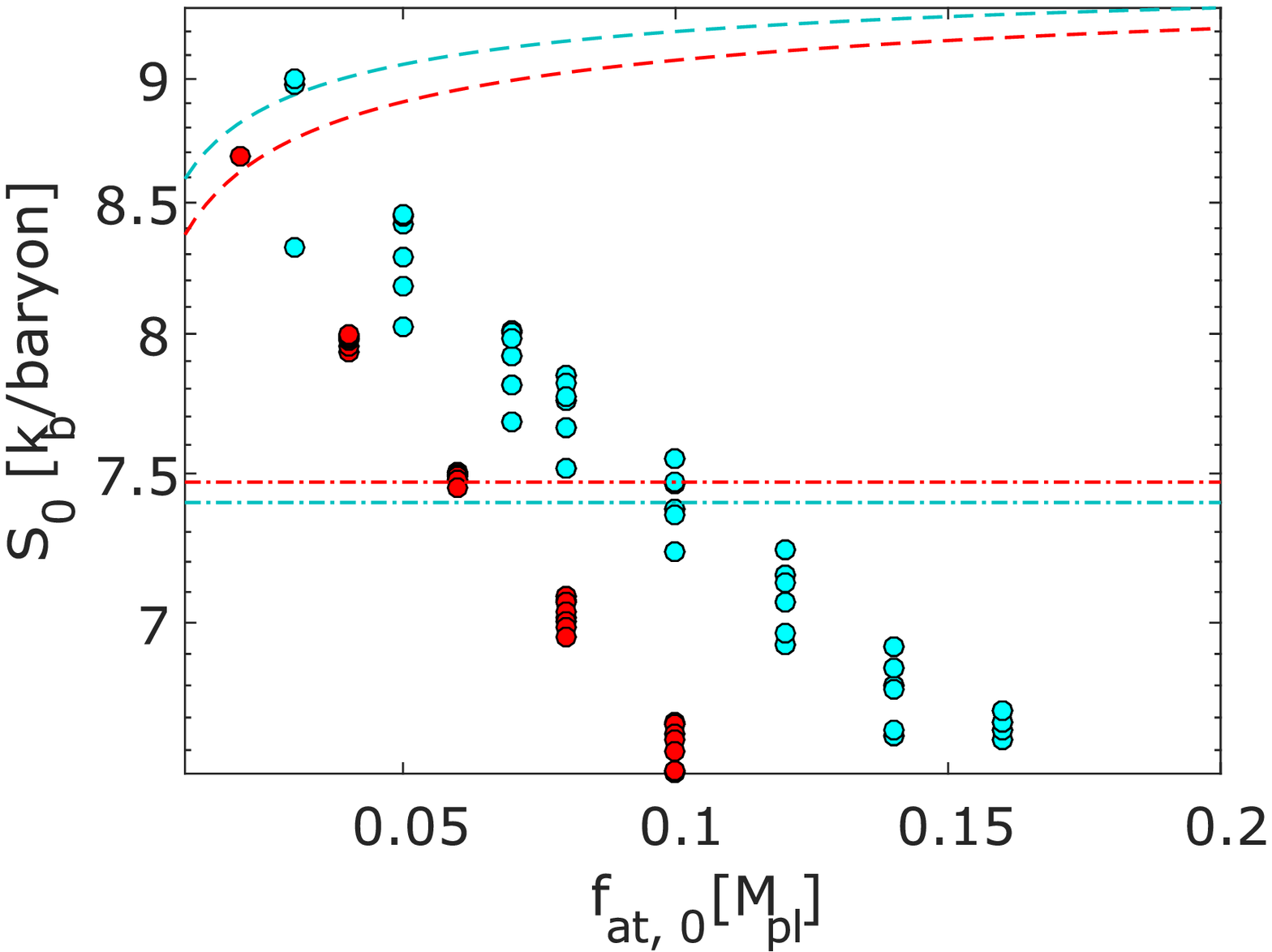}\\
  \includegraphics[width=\hsize]{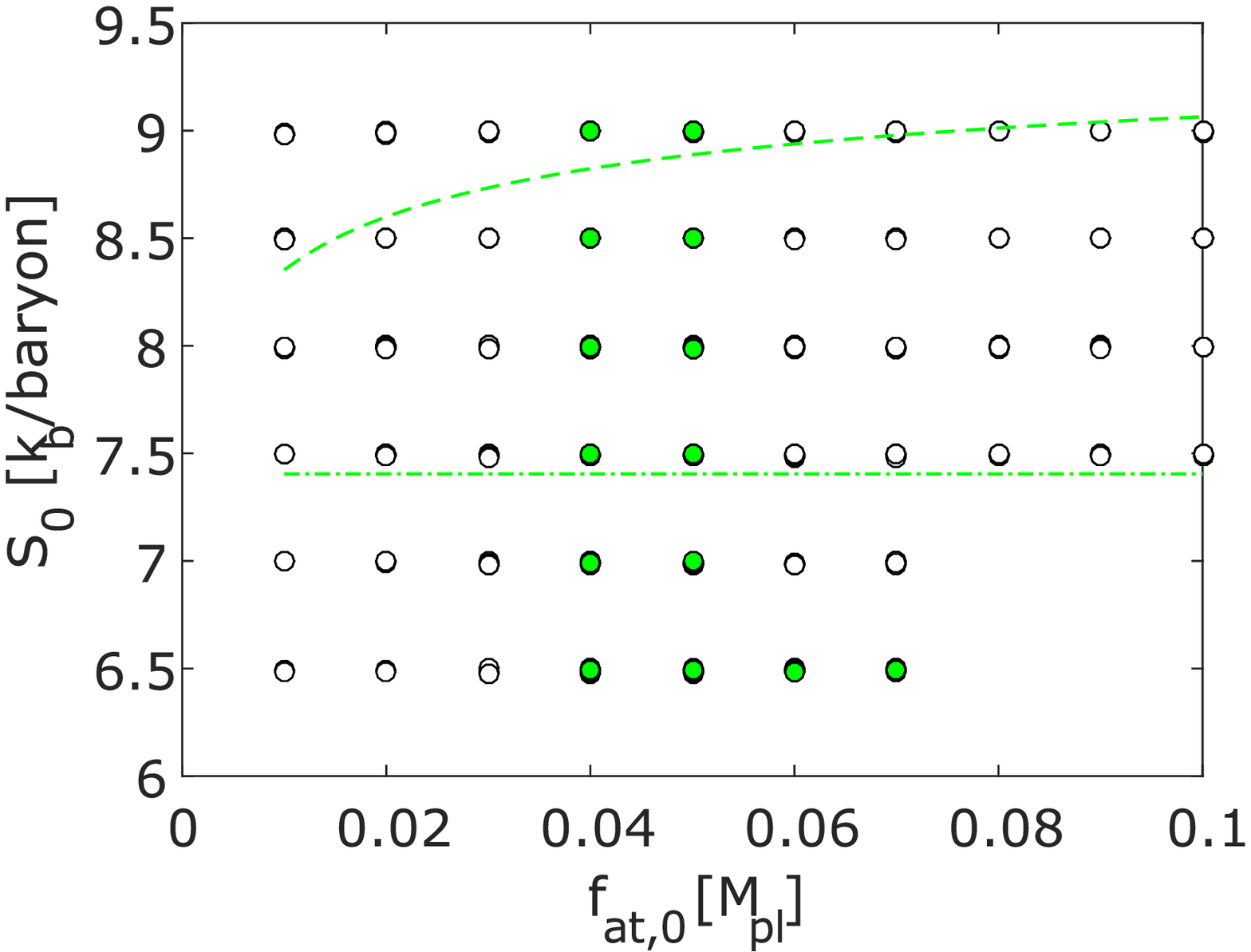}\\
  \caption{Relation between the initial atmospheric mass fractions and the central entropy of the planet for AU Mic b and K2-33 b (top planet) and Kepler-411 d (bottom planet).
  The filled circles of red, cyan and green color represent the parameters in the model that allow reproducing the observed parameters for AU Mic b, K2-33 b and Kepler-411 d, respectively.
  The black empty circles in the bottom panel correspond to all runs made for Kepler-411 d. The dashed lines show the upper limit of the estimated entropy employing the analytical approximation by \citet{mordasini2017}.
  The dashed-dotted lines show the lower limit of the entropy given by Equation 27 in \citet{malsky_rog2020}. The color of the lines is as given for corresponding planet.}\label{fig::f0-S0-young}
\end{figure}

In Figure~\ref{fig::f0-S0-young} (top panel), we show which values
of the entropy are required to reproduce the observed parameters
(i.e., the radius of the planets for a given mass, orbit, and age
of the planet, etc.) of AU Mic b and K2-33 b for a given
atmospheric mass fractions. As expected, the smaller the
atmospheric mass fraction is, the hotter the planet (and higher
the entropy) should be to match the observed radii of the planets.
We initially consider the central entropy to the interval of
6.5-9.0 ${\rm k_b/baryon}$; to further restrict this interval, we
compare our results to the predictions of two more approximations.
First, we consider the analytical approximation of the
post-formation entropy based on the formation models given in
\citet{mordasini2017}. It gives the estimation of the entropy as a
function of the absolute mass of the atmospheres; this
approximation is, however, based on the calculations focused on
hot Jupiters (up to 50~$M_{\rm jup}$). As both luminosity and
entropy grow with increasing planetary mass, we, therefore,
consider this approximation as an upper limit for the central
entropy in the planetary mass range considered here. In
Figure~\ref{fig::f0-S0-young} (dashed lines), we show the
predictions of this approximation given for different atmospheric
mass fractions assuming the mass of the planets being the upper
boundary of the observational uncertainty (and equal to 40\Mer in
case of K2-33 b). The second estimation we show here
(dashed-dotted lines) is the one suggested as the initial
luminosity by \citet{malsky_rog2020}

\begin{equation}
S_0 = 7.0 + \frac{M_{\rm pl}}{25.0 M_{\oplus}} \frac{k_b}{baryon}.
\end{equation}

By construction, this value is a lower limit of the central
entropy expected from the formation models \citep[see,
e.g.,][]{marleau_cumming2014}. We plot it in
Figure~\ref{fig::f0-S0-young} assuming for each planet the mass at
the lower limit predicted by observations (and 10\Mer in case of
K2-33 b). {We will highlight further the interval of initial
entropies outlined by these two estimations as the most realistic
values, but keep showing in plots the whole range of simulated
$S_0$ to help illustrate the general trends.}

\begin{figure}
  \includegraphics[width=\hsize]{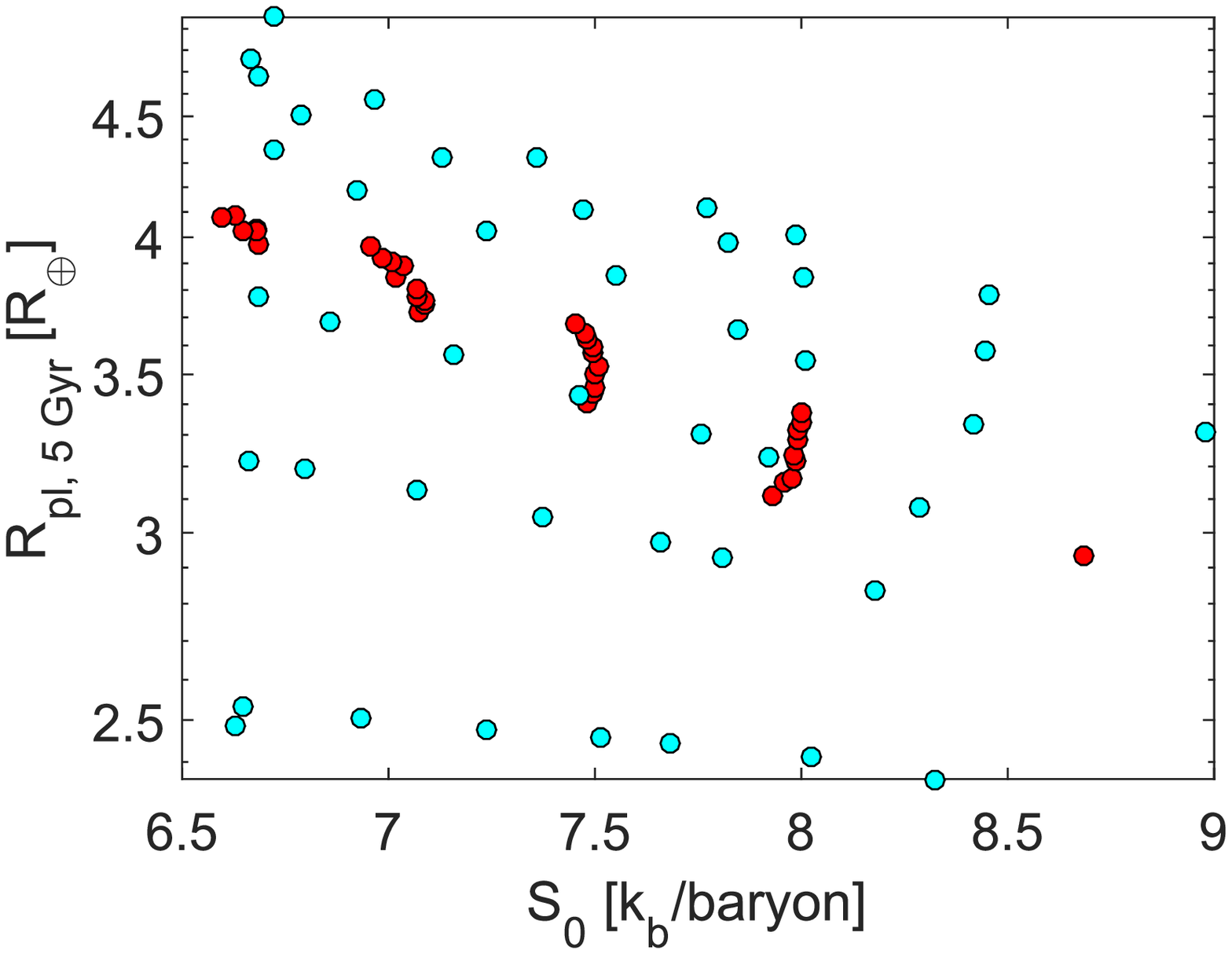}\\
  \includegraphics[width=\hsize]{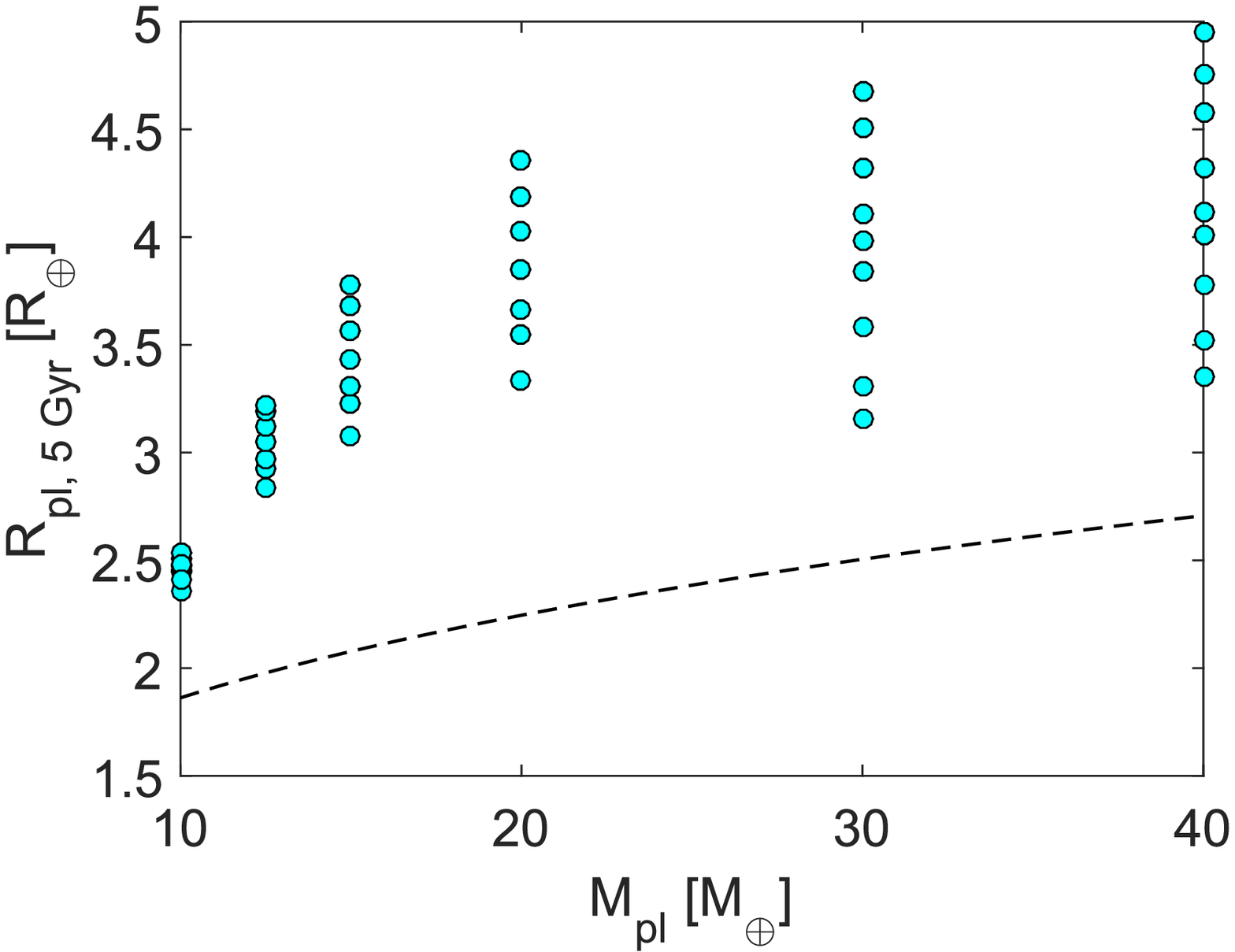}\\
  \caption{Top panel: the distribution of the planetary radii at 5~Gyr against the initial entropy of the planet. Red and cyan circles correspond to the cases of AU Mic b and K2-33 b, respectively.
 Bottom panel: the dependence of the final radius of the planet on the total mass of the planet in the case of K2-33 b. The dashed black line shows the core radius adopted for the given mass of the planet \citep{rogers2011core}.}\label{fig::S0-Rfin}
\end{figure}

After setting up the model planets reproducing parameters of AU
Mic b and K2-33 b, we evolve them for 5~Gyr to see how the choice
of the initial entropy could affect the physical properties of an
evolved planet. The choice of the final age is of minor importance
here, as all the major changes to the planet's state occur during
the first Gyr of their lifetime. In Figure~\ref{fig::S0-Rfin} (top
panel), we show the distribution of the final radii against the
initial entropy of the planet. It appears that the radii of the
evolved planets show a clear correlation with initial entropy.
This effect, however, as we will demonstrate below, is a
consequence of the degeneracy between the initial entropy and the
atmospheric mass fraction (Figure~\ref{fig::f0-S0-young}, top
panel). Therefore, for both planets, the final radii correlate
with the initial atmospheric mass fractions.

%---
This dependence is rather narrow in the case of AU Mic b due to
the relatively fine mass constraint and low atmospheric escape: it
loses no more than $\sim$4\% of its atmosphere for the lowest
planetary mass considered. As the radius of a planet depends on
the atmospheric mass fraction rather than the total mass of a
planet \citep[see, e.g.,][]{lopez2014,kubyshkina2020mesa}, the
final radius, in this case, is defined by the initial atmospheric
mass fraction and the cooling path of the planet adjusted by its
mass. We can, therefore, predict that the radius of the evolved
AU~Mic~b will be in the range of $\sim 2.9-4.1$\Rer, which implies
that the planet will safely remain in the category of
sub-Neptune-like planets. The constraint becomes more narrow if we
consider only the `realistic' values of the initial entropy ($\sim
7.5 - 9 {\rm k_b/baryon}$, as described above), giving the range
of radii of $\sim2.9-3.6$\Rer. Better constraint on mass, however,
would not improve this estimate significantly: at 5~Gyr  we can
see only a weak correlation between the radius and the mass of the
planet.

In the case of K2-33 b, however, this dependence is largely spread
due to the larger planetary mass interval, and, more importantly,
the consequent wide range of atmospheric escape rates. K2-33 is an
active star, which means that only massive enough planets can keep
most of their envelope throughout the evolution at the orbit of
K2-33~b. For the considered range of planetary masses, this is the
case for mass bins of 30 and 40~$M_{\oplus}$, which lose a maximum
of about 2.5 and 0.8\% of their envelopes. At the low-mass end of
our considered model planets, however, the atmospheric mass loss
can reach about 90\% of the initial envelope. This creates a wide
spread in radii of evolved planets between $\sim$2.4 and 6.7~\Rer.
In the top panel of Figure~\ref{fig::S0-Rfin}, the points
corresponding to the final radii of K2-33~b group into 6 distinct
lines corresponding to 6 mass bins we considered (10, 12.5, 15,
20, 30, and 40~\Mer), with the distribution of lines getting
denser at the larger radii (hence larger masses). This suggests
that to put a tight constraint on the future evolution of the
young planet orbiting an active star, one needs a good
observational constraint on the planetary mass. In the bottom
panel of Figure~\ref{fig::S0-Rfin}, we show how the radius at
5~Gyr depends on the mass of the planet. The predicted radius
grows steeply with the mass in the mass range below $\sim$30~\Rer,
and its growth slows down above this mass. Therefore, extending
the mass range considered here to heavier planets would not change
the overall result essentially.

%---

The third planet we consider here, Kepler-411 d, is about 10 times
older than AU Mic b. The age of the system is estimated to be
$212\pm31$~Myr \citep{sun2019}, which is too old for applying the
same approach as for two other planets. Based on the conclusions
made in \citet{owen2020}, we expect the post-formation luminosity
to be dissipated at this point in time. To test this statement, we
use the following scheme. We first set up an initial grid of
planets, considering planetary masses in the range of the
observational uncertainty split into 10 bins with the step of
about 1.1~$M_{\oplus}$, atmospheric mass fractions in range 1-10\%
(this range of $f_{at,0}$ allowed reproducing parameters of AU Mic
b, which has nearly the same mass) with a 1\% step, and the
initial entropies in the same range as for other planets. We then
set the initial age to 5~Myr \citep[the average lifetime of
protoplanetary disk][]{mamajek2009} and evolve each planet in this
grid for 5~Gyr, and check if the evolutionary track of planetary
radius coincides at $212\pm 31$~Myr (the age of the system) with
the measured radius of Kepler-411~d.

In Figure~\ref{fig::f0-S0-young} (bottom panel), we show the
relation between the initial atmospheric mass fraction and the
initial entropy of the planet for all model planets (empty black
circles) and mark the ones which allow reproducing parameters of
Kepler-411~d at the present time (green filled symbols). Note that
these can be reproduced only for the narrow range of initial
atmospheric mass fractions, in particular, if considering only the
`realistic' initial entropies of the planet. Given the resolution
of the grid, in the latter case only $f_{at,0}$ of 4 and 5\% are
allowed. At the same time, the radii of evolved model planets at
5~Gyr show no visible correlation with the initial entropy, and
the radii at the present age of the system show only a very weak
correlation (we do not show these plots here). This confirms the
statement formulated at the beginning of this section, that the
initial entropy of the planet has no or only minor effect on the
final radius distribution of the evolved planets \citep[which is
consistent with conclusions made in][]{owen2020}.

\begin{figure}
  \includegraphics[width=\hsize]{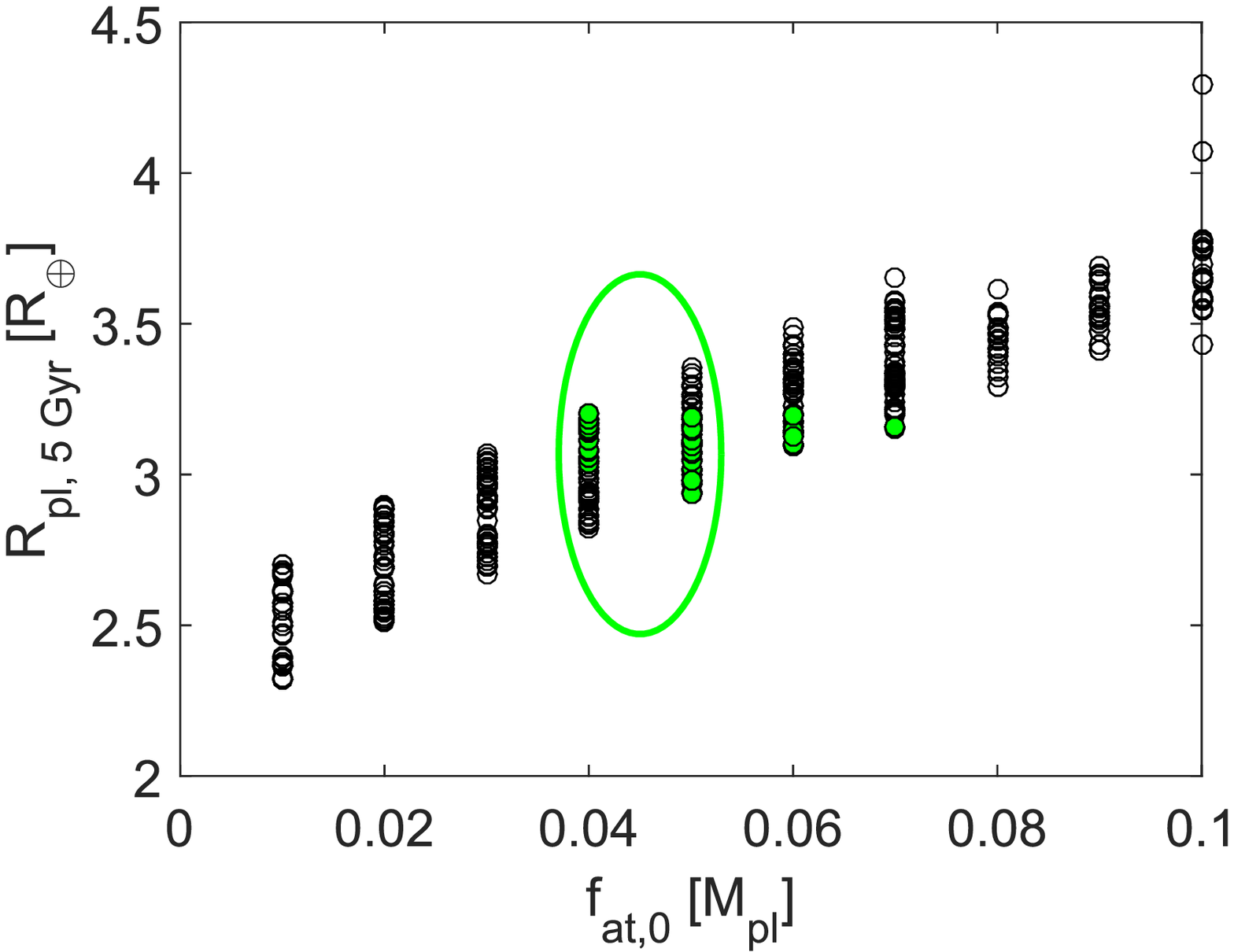}\\
  \includegraphics[width=\hsize]{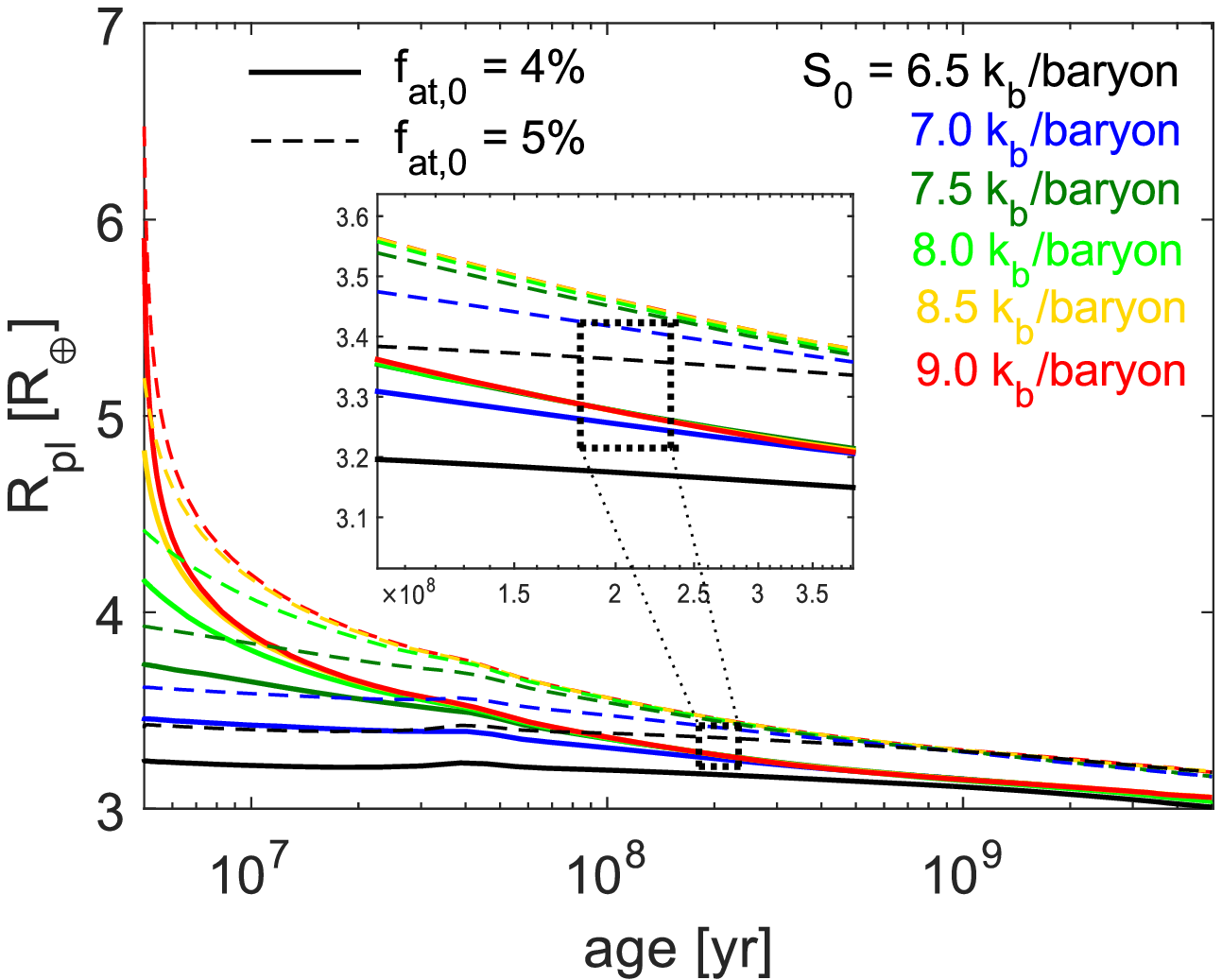}\\
  \caption{Top panel: the planetary radius distribution at the age of 5~Gyr against the initial atmospheric mass fraction. The empty black circles show all the modelled planets, and the green circles show those planets whose initial conditions allow reproducing the present time parameters of Kepler-411~d.
  The green ellipse denotes the planets in the `realistic' range of the initial entropies.
  Bottom panel: the evolutionary tracks of the radii of planets with mass equal to 15.8~$M_{\oplus}$, and initial atmospheric mass fractions of 4 (solid lines) and 5\% (dashed lines), and various initial entropies (color coded in the plot legend).
  The black rectangles denote the age and the radius measurements of Kepler-411~d. The sub-plot is the zoom-in at 100-400~Myr.}\label{fig::kepler-411d}
\end{figure}

To demonstrate the mechanism, we show in
Figure~\ref{fig::kepler-411d} (bottom panel) the evolutionary
tracks of planetary radii for planets with the mass of 15.8~\Mer
(the only mass in our grid allowing us to reproduce Kepler-411 d
with more than one $f_{at,0}$ for the `realistic' range of
entropies), initial atmospheric mass fractions of 4\% (solid
lines) and 5\% (dashed lines) and different initial entropies
(color-coded), denoting the age and radius of the planet by the
black dashed rectangle. For both atmospheric mass fractions, at
the initial stage, the various entropies of the planet create a
spread in radii of about a factor of 1.8. However, as the
heat-transfer rate is proportional to the temperature difference
with the environment ({hence, to the temperature gradient in the
atmosphere given that the temperature at the upper boundary is
fixed and is defined by the stellar irradiation}), the spread in
radii diminishes with time, and all the evolutionary tracks for
the specific initial atmospheric mass fraction converge before
1~Gyr. At the age of the system, $\sim$212~Myr, the uncertainty of
the radius measurements (which is about 3\%, i.e., rather small,
in case of Kepler-411~d) is too large to distinguish between the
different initial entropies of the planet.

On the other hand, this effect ensures that in case of the
relatively low atmospheric mass losses (which are, in the case of
Kepler-411~d, comparable to those of AU Mic b) one can use the
radius of the planet older than about 100~Myr (assuming other
parameters of the system to be known with sufficient accuracy) to
put a constraint on the initial atmospheric mass fraction of the
planet, and therefore further inform planetary formation models.
The narrow constraint on the initial atmospheric mass fraction
also allows to track the planetary evolution into the future and
provide a strong prediction on the radius of the planet at the
ages older than 1~Gyr. In the top panel of
Figure~\ref{fig::kepler-411d}, we show the dependence of the
planetary radius at 5~Gyr against the initial atmospheric mass
fraction. The designations are the same as in
Figure~\ref{fig::f0-S0-young}, i.e., the points allowing to
reproduce Kepler-411~d are shown with green circles. We
additionally highlight the points with `realistic' initial
entropies with the green ellipse. One can see, that the radius of
Kepler-411~d at 5~Gyr lies in the narrow range of
$\sim$2.9-3.2\Rer, thus, as in the case of AU Mic~b, the planet
will likely remain in the category of sub-Neptunes.

In case of strong atmospheric mass loss, both tasks discussed
above (i.e., resolving the initial atmospheric mass fraction
and/or predicting the radius of the evolved planet) will be more
complicated. The former one we have discussed in detail in
\citet{kubyshkina2019b}. The issue is that with increasing
atmospheric mass, the total mass of the planet increases only
slightly, while its radius can increase by a few times its core
radius. The atmospheric escape increases with the radius as at
least $\sim R_{pl}^3$ \citep[see, e.g.,][]{kubyshkina2018approx},
which leads to the effect visually similar to the one illustrated
in the bottom panel of Figure~\ref{fig::kepler-411d}, but
different by nature: the larger is the excess of the atmospheric
mass fraction (hence radius), the larger becomes the atmospheric
mass loss rate. Thus, the evolutionary tracks of the planets with
initial atmospheric mass fractions above a certain value will
converge with those that started with smaller atmospheres after a
few tens of Myr. Therefore, in the case of intensively escaping
atmospheres, one can often constrain only the lower boundary of
the possible initial atmospheric mass fraction from present time
parameters for planets older than 1~Gyr, or even being unable to
put any substantial constraint in case of low planetary masses.
{Among the planets simulated here, it is the case for the lower
mass bin of K2-33~b (see the lower line-shaped group of cyan
points in Figure~\ref{fig::S0-Rfin}), where the spread in the
final radius is about 7\%, which is comparable to the average
observational uncertainty for close-in low-mass
planets\footnote{According to the data from
https://exoplanetarchive.ipac.caltech.edu/ for planets with masses
smaller than 0.1~$M_{\rm jup}$ within 0.2~AU from the host star,
for which the radii are known.}.}

The second task (putting a constraint on the radius of the evolved
planet for planets as old as a few tens to a hundred Myr) becomes
also more complicated in case of strong atmospheric mass loss. As
we have shown here for K2-33 b, the strong escape leads to the
larger spread in the predicted radius distribution, and, in the
case of the poor constraint on the planetary mass, can make it
statistically insignificant.

To conclude, we have confirmed here that the initial entropies of
planets have a minor (or absent) effect on the population of the
evolved planets older than $\sim$1~Gyr, and only affects it at the
younger ages. In general, when setting the planet in MESA, the
initial central entropy is set up more or less arbitrarily
\citep[see details in][]{paxton2013}. For the range of planets
considered here, this value is $\sim 8.5$~${\rm k_b/baryon}$
throughout the grid. In our MESA models, we enforce that the
initial entropy assigned by MESA is in between the estimations
given by \citet{mordasini2017} and \citet{malsky_rog2020} (i.e.,
in the `realistic' range).

\section{Evolved population: dependencies on orbital distances and host star parameters}\label{sec::results}

To outline how a particular path of the stellar rotational (hence,
activity) evolution and the orbital semi-major axis of the
specific planet affects its final parameters, we consider here the
following grid of planets. We consider in total 6 model stars with
the masses of 0.5, 0.75, and 1.0~$M_{\odot}$, each of them
evolving as a fast or a slow rotator. The particular models were
described in Section~\ref{ssec::model-Mors}. For each star we
consider the following grid of orbital distances: 0.03, 0.04,
0.05, 0.075, 0.1, 0.15, 0.2, 0.3, and 0.5~AU. We further adjust
this grid for lower mass stars to exclude the orbits with
equilibrium temperatures of the planet lower than 300~K, as these
temperatures are not covered by the grid of hydrodynamical upper
atmosphere models \citep{kubyshkina2018grid} on which our
estimations of the atmospheric escape rates are based. This allows
only the following orbital ranges for lower mass stars:
$\leq0.15$~AU for 0.5~$M_{\odot}$, and $\leq0.3$~AU for
0.75~$M_{\odot}$. At the outer boundary of these orbital
separation ranges, however, most of the planets considered here
are going to preserve most of their primordial atmosphere, thus
their evolution is going to be dominated by purely thermal
evolution (i.e., cooling and contraction).

We consider the planetary masses of 5.0, 7.5, 10.0, 12.5, 15.0,
and 20.0~$M_{\oplus}$. These are the total masses of planets,
already including the atmospheric masses, which is done for the
convenience of comparison to the observations, where one knows the
total mass of the planet but not the atmospheric mass fraction. If
the latter is more than $\sim$10\% of the planetary mass, the mass
of the atmosphere can not be treated as insignificant relative to
the core mass anymore. In the present work, we consider the
atmospheric mass fractions of 0.5, 1, 2, 3, 5, 7, 10, 20, and
30\%.

\subsection{Overview of the results}\label{ssec::results-overview}

\begin{figure*}
  \includegraphics[width=0.49\hsize]{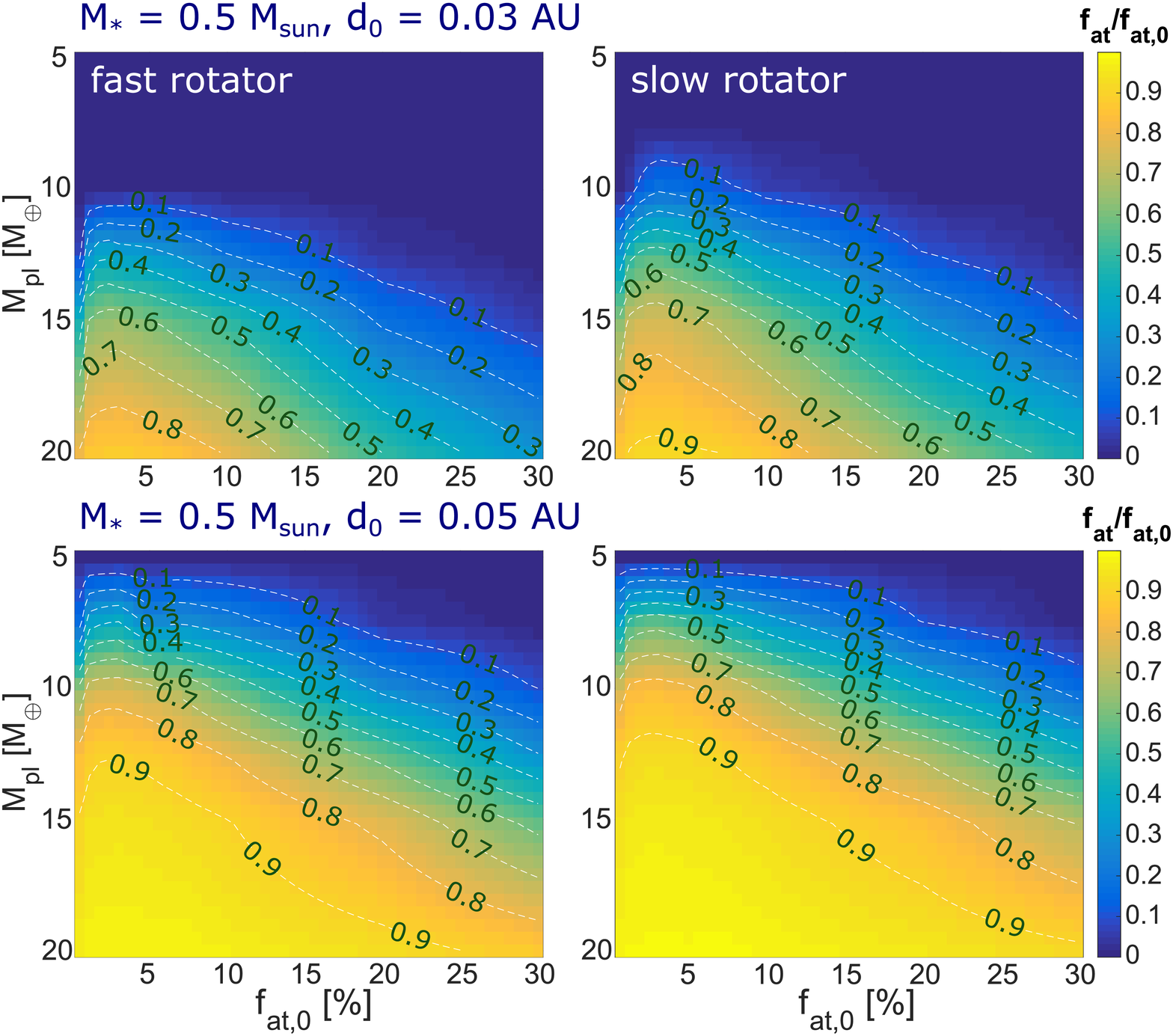} \includegraphics[width=0.49\hsize]{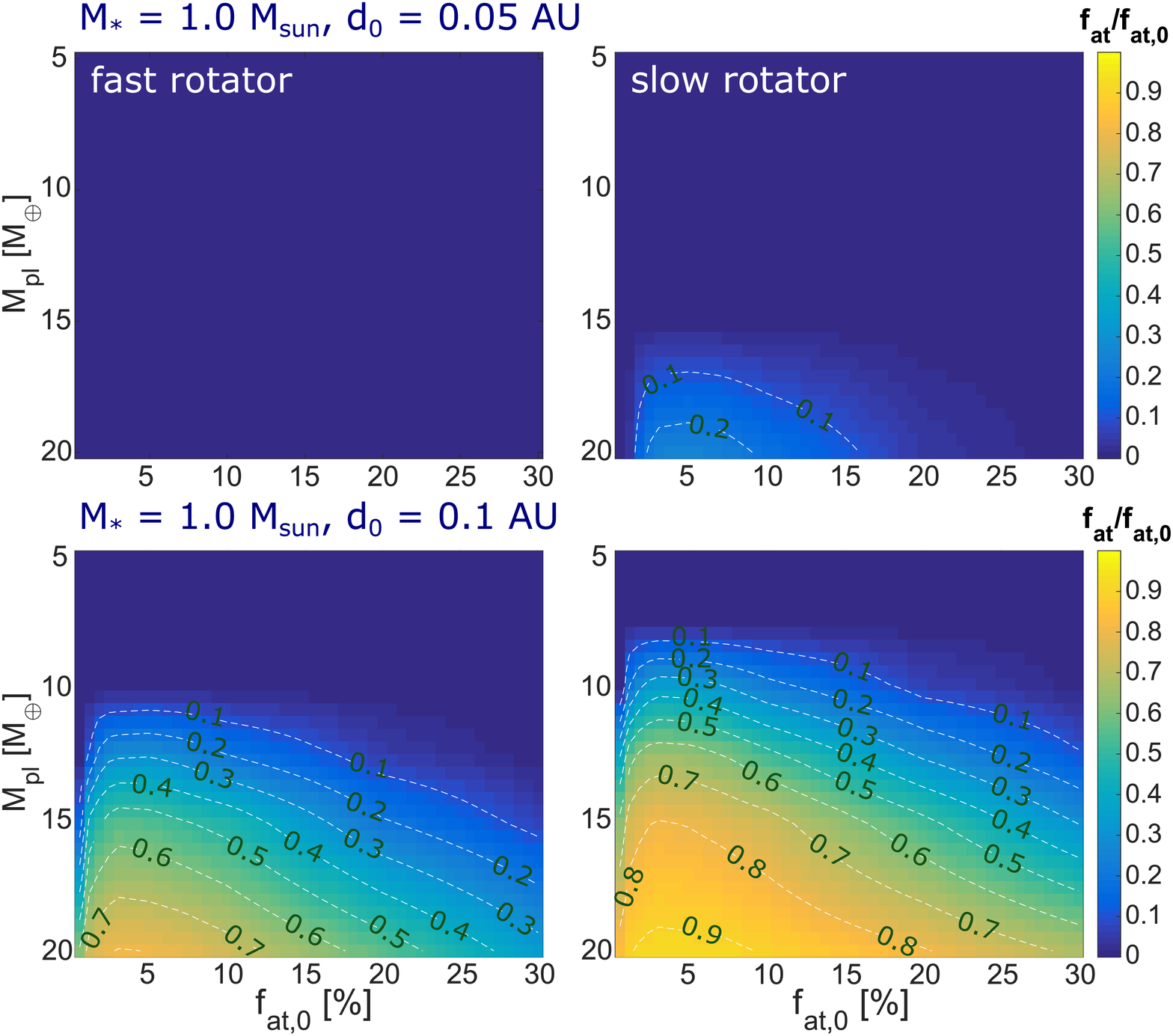}\\
  \caption{The relation of the survived atmosphere after 5~Gyr of evolution to the initial atmosphere for planets around 0.5\,\Msun star evolving as fast (first column) and slow (second column) rotator, and 1.0\,\Msun star, also as fast (third column) and slow (fourth column) rotators.
  For each star, two orbital separations are shown: 0.03 (top) and 0.05~AU (bottom) for 0.5~\Msun star, and 0.05 (top) and 0.1~AU (bottom) for 1.0~\Msun star.}\label{fig::fat2fat0}
\end{figure*}

In Figure~\ref{fig::fat2fat0} we show the ratio of the atmospheric
mass that survived over the 5~Gyr of evolution as a function of
planetary mass and the initial atmospheric mass fraction. {The
distributions of atmospheric mass fractions and planetary radii at
5~Gyr are available for each orbital separation and each stellar
model described above and can be found in
Appendixes~\ref{apx::fat} and \ref{apx::rfin}.}  To outline the
main features, here we show the distributions for planets orbiting
the 0.5~\Msun star evolving as fast (first column) or slow (second
column) rotator at 0.03 and 0.05~AU, and for planets orbiting the
solar mass star, fast (third column) or slow (fourth column)
rotator, at 0.05 and 0.1~AU. We can see that the general behavior
is similar in all cases: the amount of the atmosphere that
survived the evaporation throughout the evolution increases
towards larger planetary masses, larger orbital distances, and
lower stellar activity (hence larger initial stellar rotation
periods); i.e., the heavy planets receiving smaller amounts of
stellar radiation are less subject to atmospheric escape.

As was demonstrated in, e.g., \citet{Chen_rog2016},
\citet{kubyshkina2020mesa} for a certain orbital separation
(0.1~AU), the planets starting their evolution with relatively
compact envelopes preserve a larger fraction of their initial
atmospheres throughout the evolution. In
\citet{kubyshkina2020mesa}, the optimal initial atmospheric mass
fraction was found to be in between 3-5\% of the total mass of the
planet for the whole planetary mass range (same as in the present
paper). {In Figure~\ref{fig::fat2fat0}, this effect can be clearly
seen by the white contour lines: for the planet of specific mass
(y-axis) the relation $f_{\rm at}/f_{\rm at,0}$ maximizes in a
region of $f_{at,0}$ where the line peaks.} One can see that this
effect remains at different orbits around different stars;
moreover, the value of the optimal initial atmospheric mass
fraction remains approximately the same everywhere, with only the
maximum fraction of the preserved atmosphere changing according to
the higher/lower atmospheric escape rates for a specific star and
orbit. The effect diminishes at large orbital separations, where
the majority of the planets considered here preserve most of their
atmospheres. We will return to the discussion of this effect in
the next section.

In Figure~\ref{fig::fat2fat0}, one can see that at the same
orbital separation (see 0.05~AU), the larger fraction of the
atmosphere survives for planets orbiting lower mass star rather
than higher mass star due to the lower XUV flux received (see
Figure~\ref{fig::mors_tracks}) and lower equilibrium temperature.
The simulated planets orbiting the solar mass star at 0.05~AU (the
third and the fourth panels in the top row) essentially lose all
of their atmospheres, while at the same separation around the star
half as heavy as Sun (the first and the second bottom panels) they
preserve up to 90\% of their envelopes.

\begin{figure*}
  \includegraphics[width=\hsize]{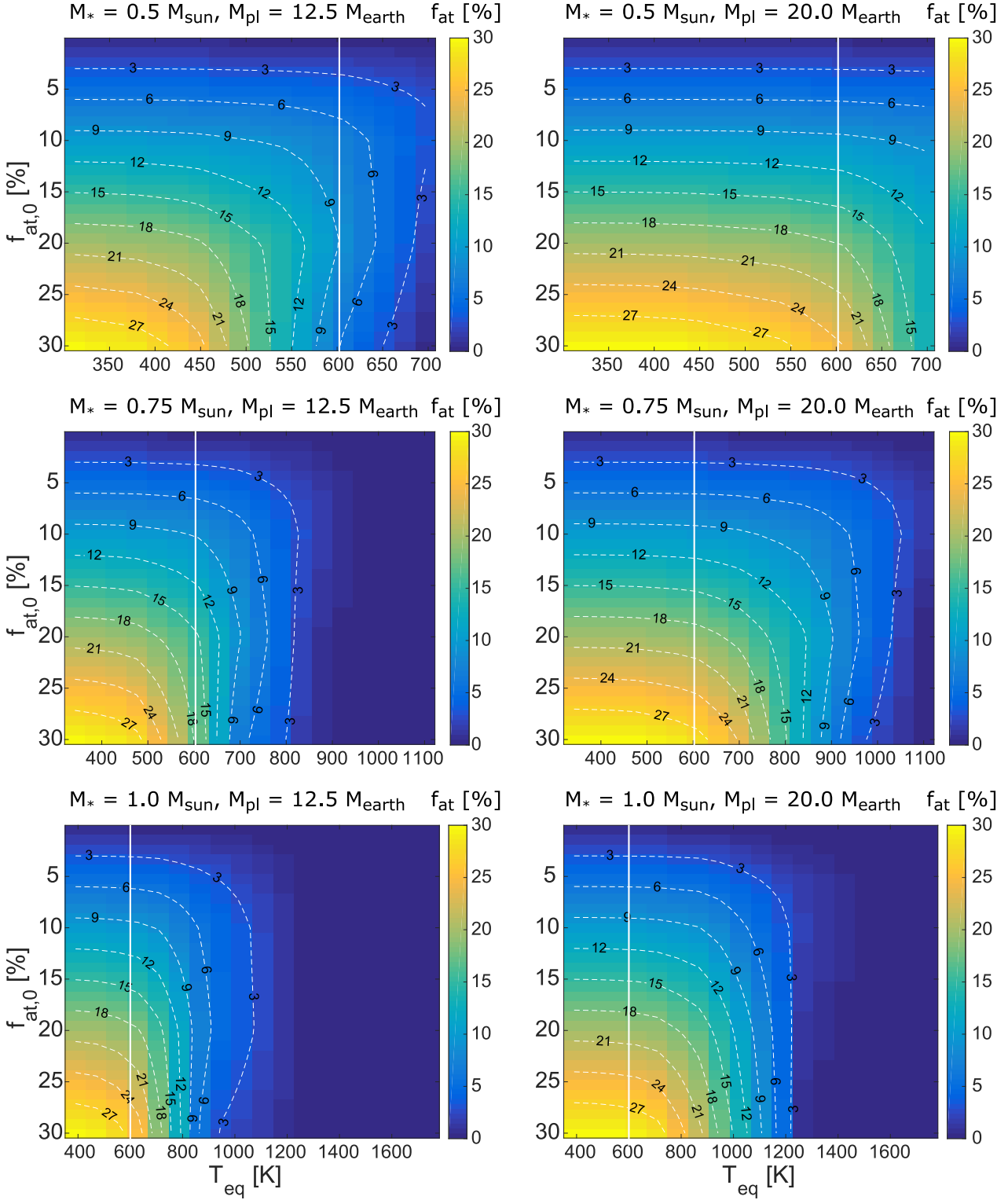}\\
  \caption{The absolute atmospheric mass fraction against the equilibrium temperature and initial atmospheric mass fraction for the slowly rotating stars ($P_{rot}(150 Myr) = 7.0$~days) of 0.5~\Msun (top row),
  0.75~\Msun (middle row), and 1.0~\Msun (bottom row), and two planets of 12.5~\Mer (left column) and 20.0~\Mer (right column).}\label{fig::fat-teq}
\end{figure*}

When it comes to habitability studies, however, more reasonable is
to compare the planets having the same equilibrium temperature
rather than the same orbital distance. By definition

\begin{equation}\label{eq::Teq}
    T_{\rm eq} = (\frac{L_{\rm bol}(1-A_{\rm b})}{16\pi\sigma_{\rm SB} d_{0}^{2}})^{1/4},
\end{equation}

\noindent where $A_{\rm b}$ is the albedo of the planet assumed to
be 0 in this study, and $\sigma_{\rm SB}$ is the Stephan-Boltzmann
constant. In Figure~\ref{fig::fat-teq} we show the absolute values
of the planetary atmospheric mass fractions at the age of 5~Gyr
against the initial atmospheric mass fraction (y-axis) and the
equilibrium temperature (x-axis). The temperature values are given
by the stellar models \citep{Johnstone2020mors,Spada2013} at a
given age. We show the distributions for two planets with masses
of 12.5 and 20~\Mer (left and right columns, respectively), and
stars of 0.5, 0.75, and 1.0~\Msun (rows from top to bottom)
evolving as slow rotators. As the temperature ranges covered for
different stellar masses are not the same, we mark the equilibrium
temperature of 600~K with a white line throughout the panels for
easy comparison.

The atmospheric mass fraction forms a $\pi$-shaped 2D
distribution, where the right edge is shaped by the atmospheric
escape. The position of this edge is, as expected, not the same
for different planetary masses, and also not the same for
different stellar masses. For the 12.5~\Mer-planet, the position
of the edge corresponding to $\sim$3\% atmospheric mass fraction
moves from about 650~K at the 0.5~\Msun star to about 900~K at the
1.0~\Msun star. That means that, at the same equilibrium
temperature, a planet around a more massive star preserves more of
its atmosphere, than around the low-mass star. {This is due to the
larger orbital separation}, hence lower $F_{\rm XUV}$ level
despite the larger stellar activity. Though being trivial, this
fact is often overseen when discussing the habitability of M-stars
\citep[][]{Johnstone2020mors}.

\subsection{Survival timescales of atmospheres: lucky parameters for prohibited areas}\label{ssec::results-tau}

As we have discussed in \ref{ssec::results-overview}, the planets
with different initial atmospheric mass fraction have different
probabilities of keeping their primordial envelope throughout the
evolution. Considering the relative amount of the atmosphere
(i.e., $f_{\rm at}/f_{\rm at,0}$, Figure~\ref{fig::fat2fat0})
preserved after Gyrs, the optimal initial atmospheric mass
fraction lies around 3-5\% of the planetary mass for all simulated
planets and all orbital separations and XUV levels. This can be
seen by the shape of the isocontours, which peak at this region of
$f_{\rm at,0}$.

\begin{figure}
  \includegraphics[width=\hsize]{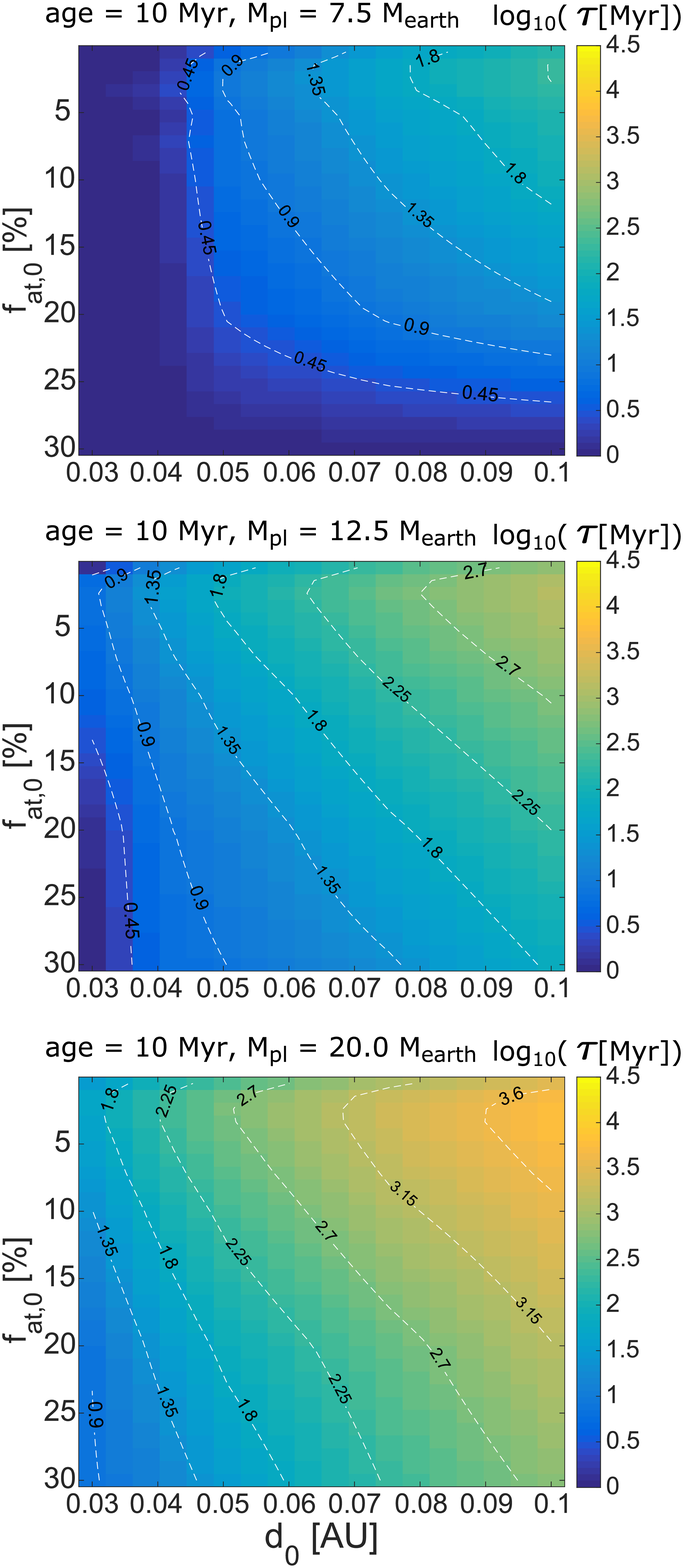}\\
  \caption{The survival timescale of the atmospheres at the age of 10~Myr for planets orbiting 1.0~\Msun star evolving as a fast rotator ($P_{rot}(150 Myr) = 1.0$~days) against the orbital separation and initial atmospheric mass fraction.
  Planetary masses are 7.5 (top panel), 12.5 (middle panel), and 20~\Mer (bottom panel).}\label{fig::tau}
\end{figure}

Another parameter allowing to quantify this effect is the
atmospheric survival timescale $\tau = M_{\rm atm}/{\dot{M}}$ at
the early ages, where $M_{\rm atm}$ is the current mass of the
atmosphere and $\dot{M}$ is the atmospheric escape rate at the
same moment. In Figure~\ref{fig::tau} we show $\tau$ at the age of
10~Myr for planets with masses of 7.5, 12.5, and 20~\Mer orbiting
the solar mass star evolving as a fast rotator at various orbits.
We restrict our consideration to the orbital distances smaller
than 0.1~AU, as this region is the most affected by atmospheric
evaporation, and for most planets in our sample, the border
between lost/preserved atmospheres lies within it. In terms of the
atmospheric survival timescale at 10~Myr, the lost/preserved
border is at $\tau\sim$100-200~Myr, which roughly corresponds to
the 2.25 line in Figure~\ref{fig::tau} (in logarithmic scale): for
a given planetary mass, planets to the right of this line (towards
orange/yellow contours) will certainly keep some of their
atmospheres.

In line with what was shown before for relative atmospheric mass
fractions of the evolved planets in
Section~\ref{ssec::results-overview}, the dependence of the
atmospheric survival timescale on the initial atmospheric mass
fraction is not monotonic: $\tau$ maximizes at the value of
$f_{\rm at}\sim 3\%$. The shape of $\tau(f_{\rm at,0})$,
therefore, suggests that at certain close-in orbital separations a
planet of specific mass would be capable of keeping part of its
atmosphere only for a restricted interval of initial atmospheric
mass fractions (see, e.g., the 0.07~AU for 12.5~\Mer planet or
0.05~AU for 20~\Mer planet in Figure~\ref{fig::tau}). Thus, at
5~Gyr, the \emph{absolute} value of the atmospheric mass fraction
of 12.5~$M_{\oplus}$-planet at 0.07~AU will be non-zero only if it
started with $f_{\rm at,0}$ in range of $\sim 1-7\%$.

This might seem contradictory at first. For the small initial
atmospheres, this effect is clear -- smaller atmospheres are lost
faster. For the larger atmospheres, the radius of the planet grows
with $f_{\rm at}$ and consistently grows the escape rate, which
leads to larger atmospheres escaping faster at the early stages of
evolution. However, one would expect that after reaching the same
amount of the atmosphere as the planet that started with smaller
$f_{\rm at,0}$, the planet with the larger initial envelope would
proceed along the same path (similar to the effect shown in
Figure~\ref{fig::kepler-411d}, bottom panel, for the (nearly pure)
thermal evolution of Kepler-411~d). Thus two planets that started
with compact and extended atmospheres are naively expected to end
up having the same atmospheric mass fraction after a few Gyrs of
evolution. In the present framework, we find, however, that the
planets with larger initial atmospheric mass fractions under some
conditions end up having smaller atmospheres than those that
started with compact ones.

\begin{figure*}
  \includegraphics[width=0.49\hsize]{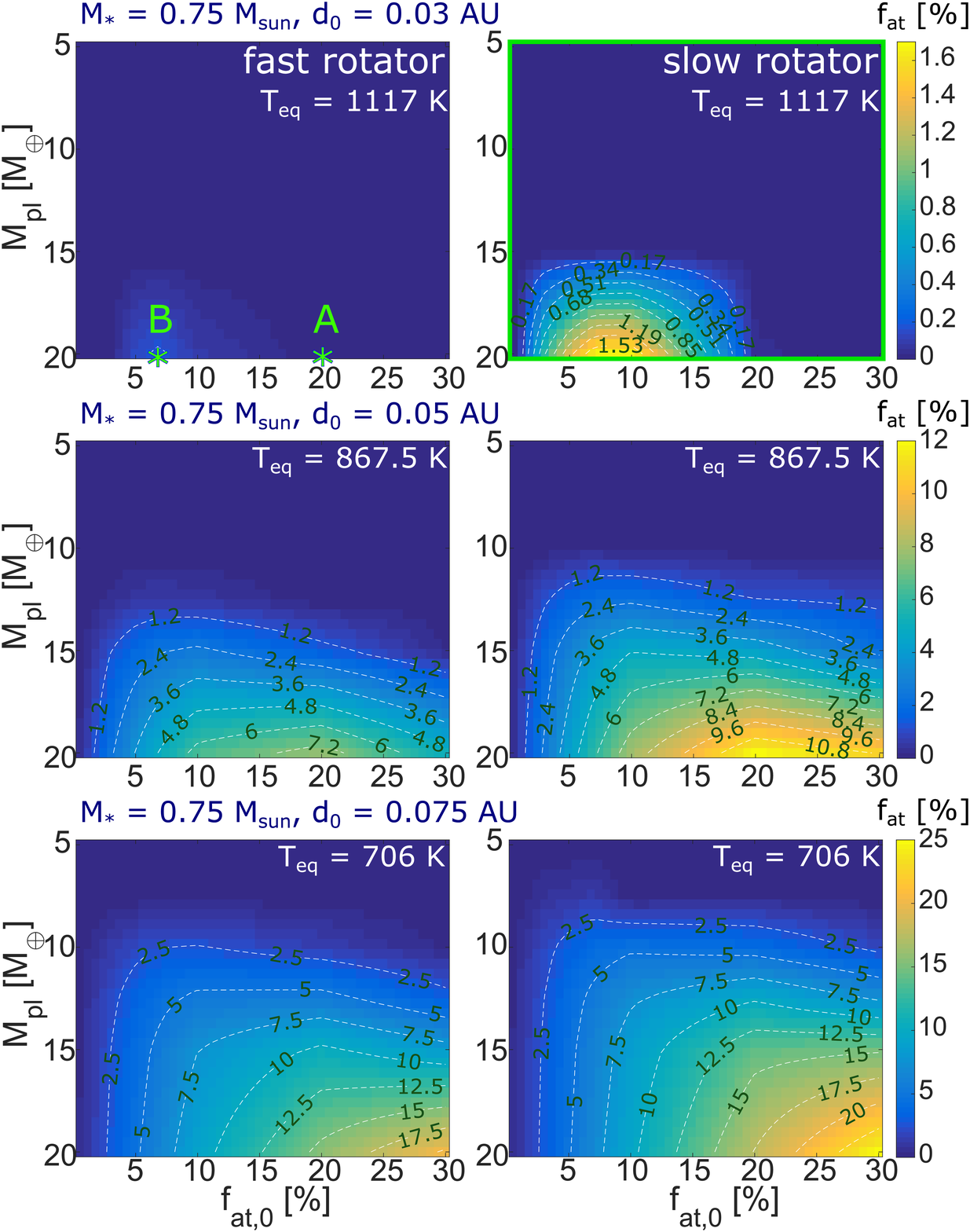} \includegraphics[width=0.49\hsize]{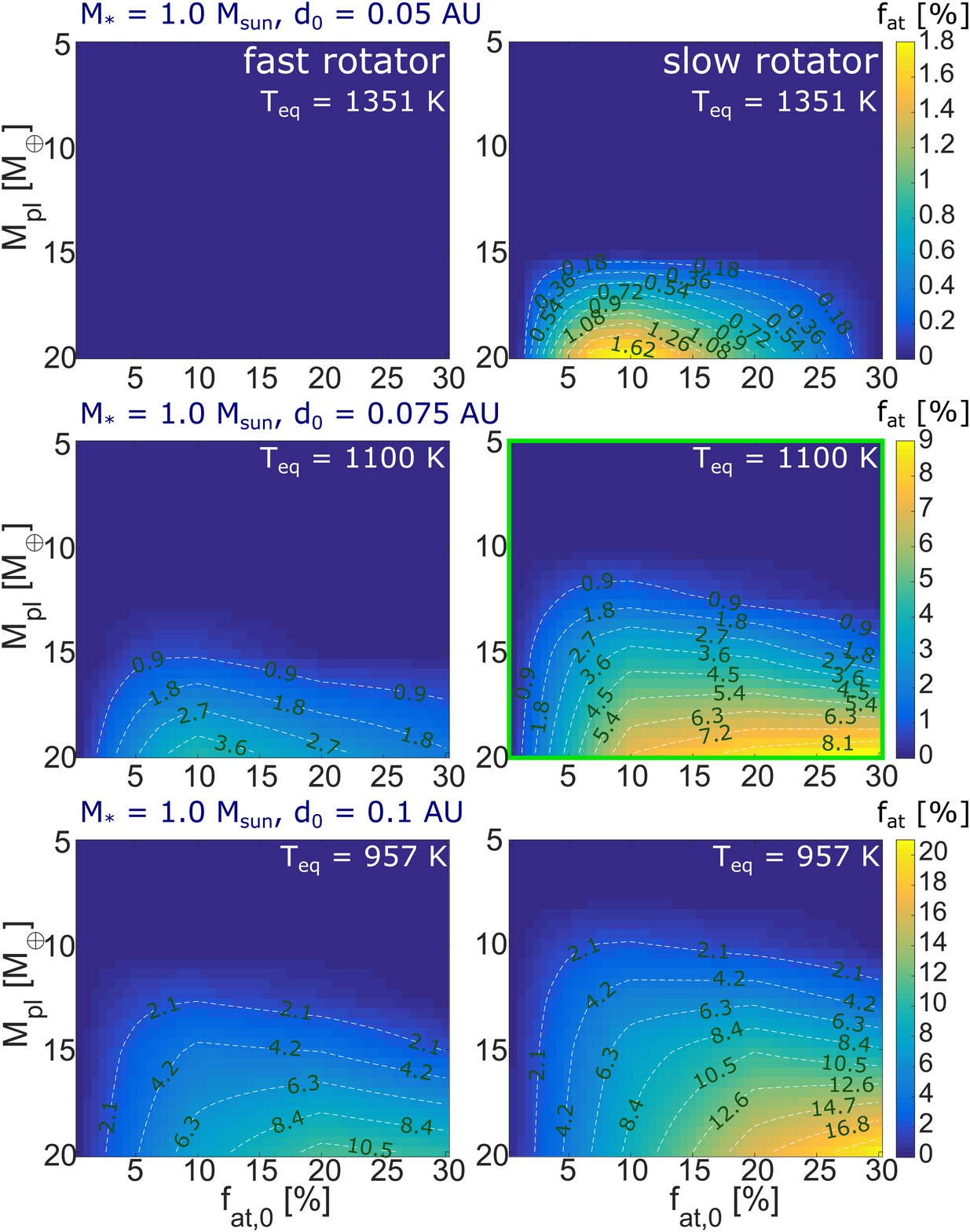}\\
  \caption{The atmospheric mass fractions of the planets at 5~Gyr around different stars against planetary masses and initial atmospheric mass fraction. Different columns correspond to different stellar models.
  First column: $M_* = 0.75$~$M_{\odot}$, fast rotator ($P_{rot}(150 Myr) = 0.5$~days). Second column: $M_* = 0.75$~$M_{\odot}$, slow rotator ($P_{rot}(150 Myr) = 7.0$~days).
  Third column: $M_* = 1.0$~$M_{\odot}$, fast rotator ($P_{rot}(150 Myr) = 1.0$~days). Fourth column: $M_* = 1.0$~$M_{\odot}$, slow rotator ($P_{rot}(150 Myr) = 7.0$~days).
  Indicative equilibrium temperatures for these stellar masses and orbital separations are indicated in each panel.}\label{fig::fat_AB}
\end{figure*}

In Figure~\ref{fig::fat_AB}, we show the atmospheric mass
fractions in \% of the total planetary mass against the planetary
mass and the initial atmospheric mass fraction for planets
orbiting four different stars at three different orbital
separations for each of them. These are the stars of 0.75 (the two
left columns) and 1.0~\Msun (the two right columns) evolving as
fast (the first and the third columns) and slow rotators (the
second and the fourth), and orbital separations of 0.03, 0.05, and
0.075~AU for 0.75~\Msun star (rows from top to bottom), and 0.05,
0.075, and 0.1~AU for the solar mass star. {In general, more
massive planets with smaller initial atmospheric mass fractions
tend to retain most of their atmospheres (orange-yellow contours).
Note also that initial atmospheres are better preserved for
planets orbiting at further out distances and around slowly
rotating stars.} One can see that, as predicted by $\tau$
distributions, at the closest-in orbits only a narrow region of
initial atmospheric mass fractions allows to keep the atmosphere
around the planet of a given mass. With increasing orbital
separation, this region moves towards the larger initial
atmospheric mass fractions and becomes broader, and finally, the
effect diminishes at the large orbital separations (at least for
the range of $f_{\rm at,0}$ considered here).

%---
{Between the cases of the fast and slow rotating star, the
quantitative difference is quite large (about an order of
magnitude at the closest in orbits and up to about twice for the
larger separations). Qualitatively, though, the distributions look
very similar, i.e., $f_{at}$ maximizes in the same region, and the
range of ``lucky'' $f_{at,0}$ does not change much (it might not
look so at closest-in separations, but it is due to the maximum
$f_{\rm at}$ being close to zero in the case of fast rotating
star). Thus, despite the effect of atmospheres surviving at
certain initial atmospheric mass fractions is clearly caused by
the atmospheric mass loss, the shape of the distribution (i.e.,
the concrete interval of $f_{\rm at,0}$ for a given mass of the
planet) is to some extent controlled by the thermal effects.}

To compare these distributions at different host stars, but at
similar equilibrium temperatures, we highlight the two panels with
orbital separations corresponding to the similar temperatures
($\sim 1100$~K) at 0.75 and 1.0~\Msun stars with the green frames.
These are 0.03~AU for 0.75~\Msun star, and 0.075~AU for 1.0~\Msun
star (Figure~\ref{fig::fat_AB}). As was shown before in
Figure~\ref{fig::fat-teq}, at the similar $T_{\rm eq}$, the planet
orbiting the more massive star preserves more of the initial
atmosphere in comparison to the one orbiting the less massive
star. In turn, the effect of the isolated ``lucky'' initial
atmospheric mass fractions is less pronounced. Therefore, it
exists in the very restricted range of close-in orbital
separations and is more relevant for low-mass stars.

\subsubsection{The source of the effect}

To understand the source of this peculiar effect we consider two
model planets as an example: both orbiting fast rotating
0.75~\Msun star at 0.03~AU orbital separation, and starting their
evolution as 20~\Mer planet with 20\% (planet A) and 7\% (planet
B) of the mass in the atmosphere (see Figure~\ref{fig::fat_AB},
top left panel). Planet B preserves a small fraction of the
atmosphere after 5~Gyr of the evolution, while planet A loses its
atmosphere completely. For this to happen, at the moment when two
planets achieve the same atmospheric mass fraction (as discussed
above), planet A has to have a larger atmospheric escape rate.

\begin{figure}
  \includegraphics[width=\hsize]{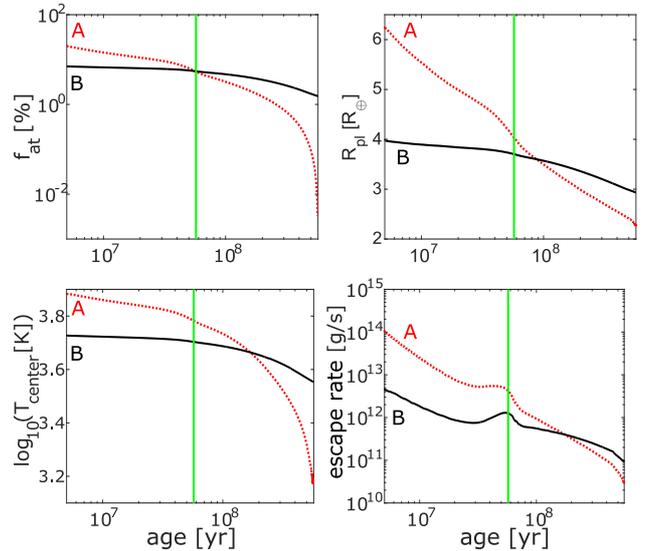}\\
  \caption{Evolutionary tracks of the 20~\Mer planet that started the evolution with 7 (black solid lines) and 20\% (red dotted lines) of its mass in the
  atmosphere. Planetary parameters are presented in the following
  order.
  Top left: the atmospheric mass fraction. Bottom left: the central (core) temperature of the planet.
  Top right: the radius of the planet. Bottom right: the atmospheric escape rate.
  The vertical green line represents the time when the atmospheric mass fractions of two planets become equal.}\label{fig::Tdiff}
\end{figure}

In Figure~\ref{fig::Tdiff}, we present the evolutionary tracks of
the atmospheric mass fraction, planetary radius, atmospheric
escape rate and the temperature of the cooling core for these two
planets. Simply from the formulation (where we include the mass of
the atmosphere into planetary mass at the beginning of the
simulation, i.e., planetary mass changes with time), it is clear
that for these two planets having the same atmospheric mass
fraction does not mean having the same total mass. Instead,
$M_{\rm pl, A} = 0.8M_0 + M_{\rm atm, A}$, and $M_{\rm pl, B} =
0.93M_0 + 1.16M_{\rm atm, A}$, where $M_0$ is the total initial
mass of the planet (which is the same for A and B), and $M_{\rm
atm, A}$ is the mass of the atmosphere of planet A. Therefore, at
the time when $f_{at, A} = f_{at, B}$ (green vertical line in
Figure~\ref{fig::Tdiff}), $M_{\rm pl, A} = 16.93$~$M_{\oplus}$,
and $M_{\rm pl, B} = 19.67$~$M_{\oplus}$. This alone is enough for
the atmospheric escape rate of planet A to be $\sim3$ times larger
than the one of planet B.

However, we noticed one further difference between the two
planets. Though two planets start their evolution having the same
core luminosity, they evolve differently through the ``disk
phase'', i.e., through the 5~Myr period that follows the
preparation stage in our simulations and precedes the main
evolution phase, when the planet cools down without atmospheric
escape. The planet with the larger atmosphere (A) cools down more
slowly than the one with the thinner atmosphere (B), and therefore
starts the evolution after the ``disk dispersal'' (the mass-loss
switches on) having a larger core temperature. After that moment,
the temperature of the core of planet A drops down faster compared
to planet B, due to the larger atmospheric escape, but not fast
enough to reach the same temperature to the point when the two
atmospheric mass fractions are equal (see bottom left panel of
Figure~\ref{fig::Tdiff}); the picture does not change
qualitatively if considering the actual mass of the atmosphere
instead of the atmospheric mass fraction. The larger temperature
results in the larger radius corresponding to the same $f_{\rm
at}$ (top right panel), which in turn increases further the
atmospheric escape rate of planet A (bottom right panel),
intensifying the effect we consider here.

To split between the `mass' and `temperature' parts of the effect,
we have reconsidered the planets in question employing the
approach where the mass of the core is treated as the main part of
the planetary mass and the atmospheric mass as a minor addition
(which is the most conventional approach and was used in
\citealt{kubyshkina2020mesa}). Thus, in the example considered
before, planets A and B start with the core masses of 20~\Mer with
the corresponding additions of the atmospheric mass on top of
this. In this formulation, at the moment when $f_{\rm at, A} =
f_{\rm at, B}$ (coming $\sim 80$~Myr later than in the example
above), the masses of planets are equal. The temperature
difference, however, remains and leads to the difference of a
factor of about two in the atmospheric escape rates. {Thus, for
planet B this difference in approach does not change the
evolutionary track a lot: it preserves slightly more of the
initial atmosphere due to the 10\% larger mass. Planet A, at the
same time, preserves its atmosphere longer (up to $\sim$1.2~Gyr
instead of 566~Myr), but is yet unable to keep it. Therefore the
same ``restricted zone'' effect can be observed also with the
alternative formulation of the problem, being caused purely by the
difference in core temperatures of the planets.}

Changing the duration of the ``disk phase'' and the initial
luminosities of planets also change the overall picture
quantitatively, but not qualitatively. However, it is important to
note that we do not consider here the actual planet formation or
the realistic protoplanetary disk phase. {As we discussed before
in Section~\ref{ssec:model-entropy}, the formation models suggest
that for different planets the post-formation luminosity (at the
time of a disk dispersal) is indeed not the same. The
approximation by \citet{mordasini2017}, which we considered in
Section~\ref{ssec:model-entropy}, predicts larger initial
luminosity for a larger mass of the atmosphere.} However, the
detailed study of the effect requires the actual application of
the formation models to set up the initial conditions of the
evolving planets and the thorough investigation of their
assumptions. This detailed investigation lies outside the scope of
the present work and is left for a future study.

In light of the discussion presented here, this effect puts some
constraints on the conclusions we made in
Section~\ref{ssec:model-entropy}, namely that the initial
luminosity does not affect the final state of the evolved planet
and therefore the overall planetary demographics. This may not be
true for planets subject to extreme atmospheric mass loss, i.e.,
for the low to intermediate mass planets at short orbits. {Having
certain initial parameters (atmospheric mass fraction and core
temperature) can affect the survival timescales of the
atmospheres, and let the planetary atmosphere survive through the
initial stage of the young and active host star. If the planet
keeps its atmosphere through the first Gyr of evolution, it is
most likely that the planet keeps it up to the average age of
5-10~Gyr.}

This also suggests that the simultaneous consideration of the
atmospheric mass loss and the thermal evolution of the atmosphere
becomes particularly important in these regions (at orbits within
$\sim0.05$~AU). Additionally, the existence of a ``lucky
interval'' in the initial parameters of the planet could explain
the existence of some peculiar planets, whose parameters seem to
be unlike for their environment, as, e.g., LTT~9779~b
\citep{jenkins2020}, or TOI-132~b \citep{diaz2020}.

%%%%%%%%%%%%%%%%%%%%%%%%%%%%%%%%%%%%%%%%%%%%%%%%%%%%%%
\section{Analytic relation between planetary radii and atmospheric mass fraction}\label{sec::approximation}

The simultaneous consideration of the thermal evolution of the
atmosphere and the atmospheric escape is being important, in
particular for the planets subject to the strong escape. However,
the relation between the atmospheric mass fraction of the planet
and its radius for a specific set of planetary parameters can be
approximated with a relatively simple function. We have derived an
analytical function in \citet{kubyshkina2020mesa} for the planets
in the mass range of 5-20~\Mer orbiting a solar mass star at
0.1~AU. In the present work, we extend such parameterization to a
wider range of physical parameters, such as different stellar
masses and orbital separation. In \citet{kubyshkina2020mesa}, we
used as a basic approximation function the cubic polynomial. Here,
we employ the following function, which allows reducing the
average mean square error throughout the whole parameter space.

\begin{equation}\label{eq::fat_tanh}
f_{\rm at}(R_{\rm pl}) = C_1(R_{\rm pl}-R_{\rm
c})^{C_2}tanh(C_3(R_{\rm pl}-R_{\rm c}))
\end{equation}

\begin{figure}
  \includegraphics[width=\hsize]{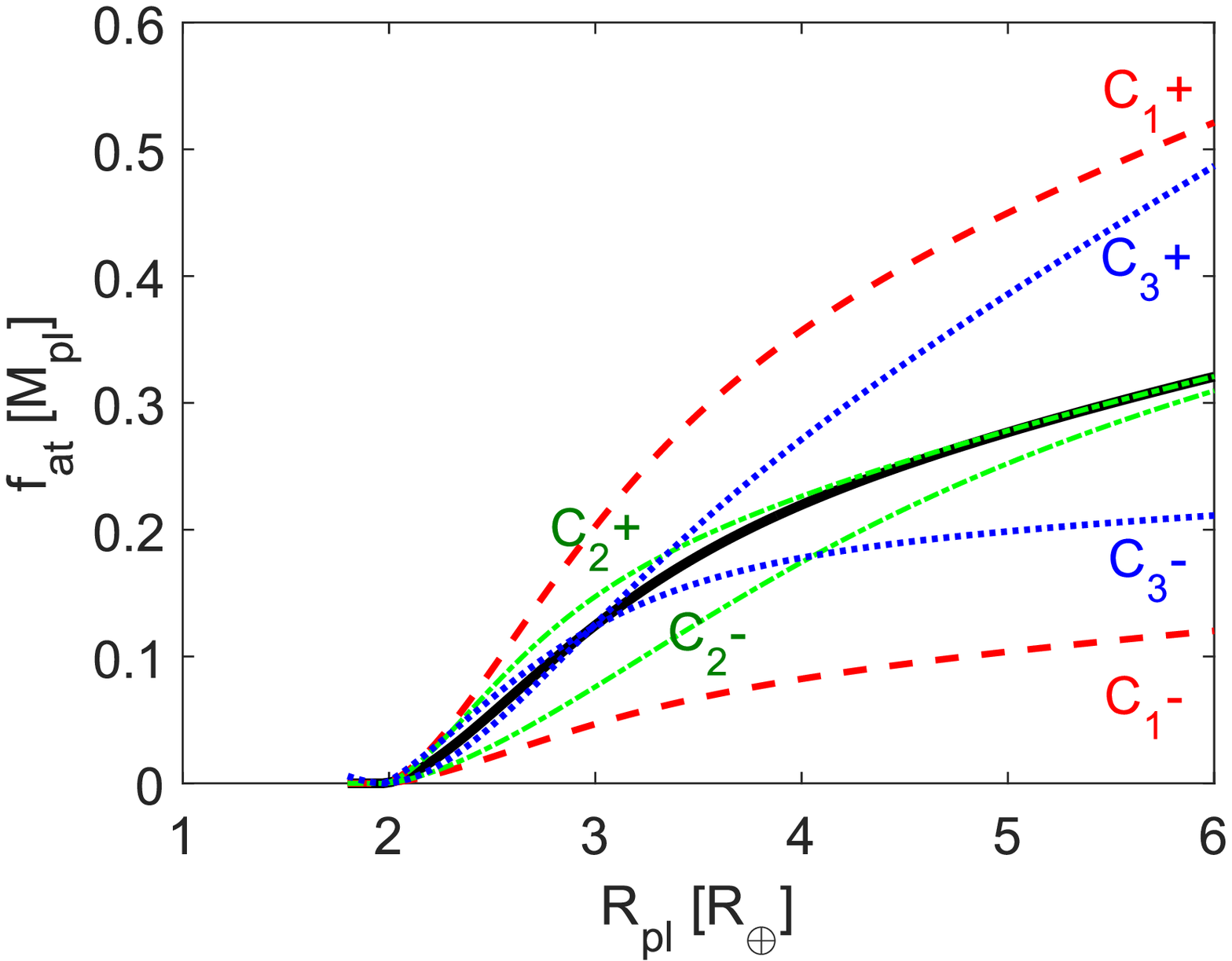}\\
  \includegraphics[width=\hsize]{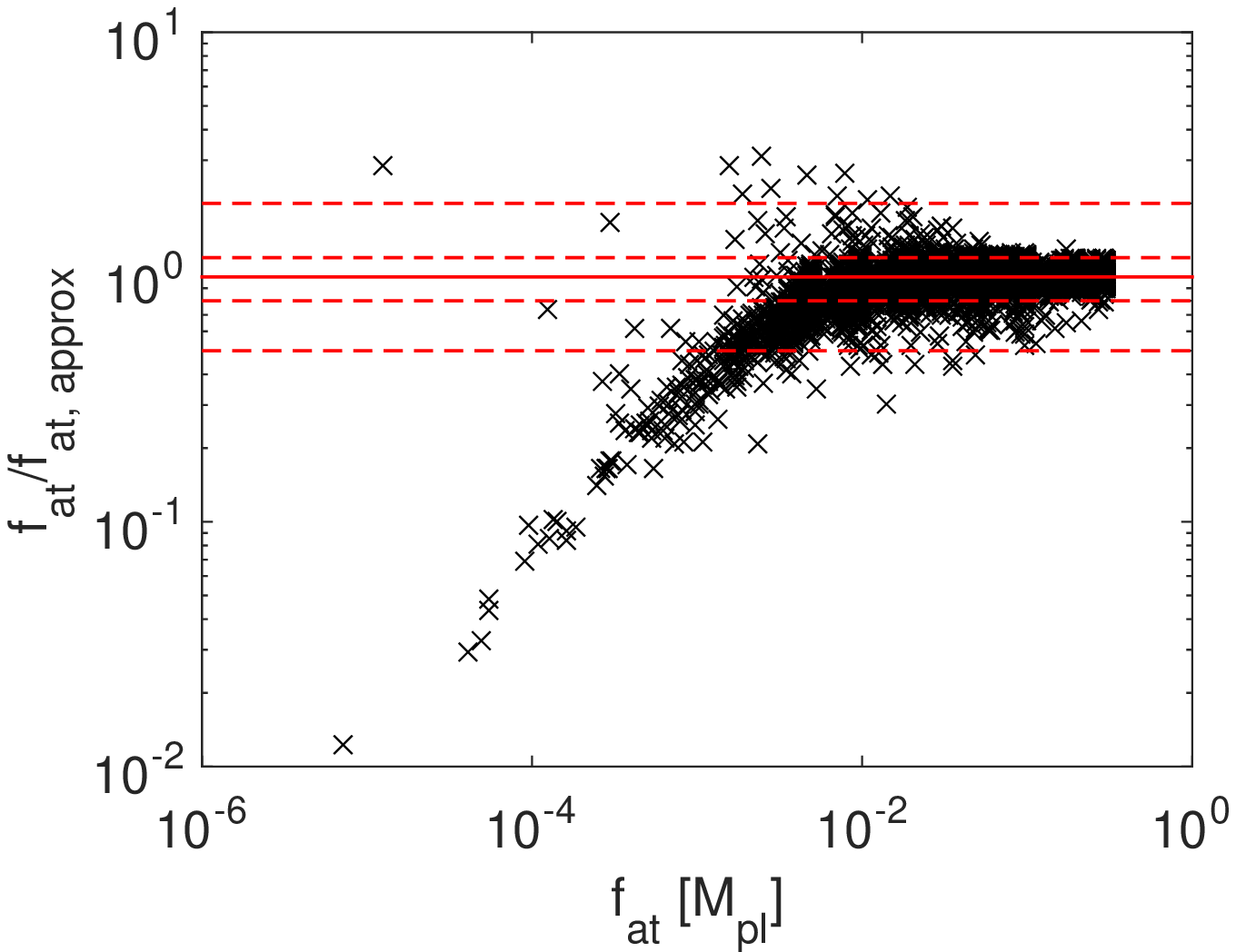}\\
  \caption{Top: the scheme presenting the typical shape of
  the $f_{\rm at}(R_{\rm pl})$ dependence given by
  Equation~\ref{eq::fat_tanh} (black solid line). The
  additional lines demonstrate how increase/decrease in coefficients $C_1$
  (top/bottom red dashed lines), $C_2$
  (top/bottom blue dotted lines), $C_3$
  (top/bottom green dashed-dotted lines) affect the shape of the
  dependence. {For clarity of the scheme, the values of $\Delta C_1$, $\Delta C_2$, and $\Delta C_3$ were taken larger than
  typical variations observed throughout the grid of simulated
  planets and are $\sim 60\%$ for $C_1$, $\sim 70\%$ for $C_2$,
  and $\sim 20\%$ for $C_3$.}
  {The core radius is assumed at 2~\Rer.}
  Bottom: To test the performance of our fit, we show the atmospheric mass fraction obtained in our MESA simulations over the value obtained from our approximation~\ref{eq::fat_tanh}, as a function of the atmospheric mass fraction.
  Black crosses represent the time-snapshots for individual planets at 100~Myr, 1~Gyr, and 5~Gyr. The red solid line denotes $f_{\rm at}/f_{\rm at, approx} = 1$, and four dashed lines correspond to (top to bottom) 2, 1.2, 0.8, 0.5. }\label{fig::approx}
\end{figure}

This function has as the argument $(R_{\rm pl}-R_{\rm c})$, where
$R_{\rm c}$ is the core radius of the planet, i.e., the thickness
of the atmosphere. This ensures that when $f_{\rm at} = 0$ the
radius of the planet equals its core radius\footnote{In MESA
simulations when $f_{\rm at}\rightarrow 0$ the radius of the
planet tends to the value slightly larger than $R_{\rm c}$, which
is likely connected to the complication of representing very thin
atmospheres due to the interpolation within the parameter tables.
We, therefore, focus on minimizing the error of approximation for
the larger atmospheres $f_{\rm at}\geq 0.5\%$.}. {For convenience, in approximation we considered $R_{\rm c} = M_{\rm pl}^{0.27\pm0.02}$, which describes well the core radii adopted in the MESA simulations. As a final value of the exponent we adopt 0.25, as it allows to minimize the approximation errors.} $C_1$,
$C_2$, and $C_3$ are free parameters of the approximation, which
we fit. In Figure~\ref{fig::approx} (top panel) we illustrate the
typical dependence of the function and how changes in each of the
coefficients affect its shape.

The main parameters of the evolving planets (except of the
argument and the function in Equation~\ref{eq::fat_tanh}, $R_{\rm
pl}$ and $f_{\rm at}$) are planetary mass, orbital separation,
equilibrium temperature and received XUV flux ($F_{\rm XUV}$). For
the star, these are stellar mass and XUV luminosity. All these
parameters change with the age of the system and are not
independent of each other. In particular, the equilibrium
temperature and the XUV flux received by the planet depend on the
orbital separation and the bolometric and XUV luminosities of the
star, while the luminosities depend on the age of the system and
the stellar mass, and the XUV luminosity, in turn, depends on the
$P_{\rm rot}(age)$. To rule out degeneracies we consider the
following.

The equilibrium temperature can be defined through the bolometric
luminosity as defined in Equation~\ref{eq::Teq}, i.e., it is
proportional to $\frac{L_{\rm bol}}{d_0^2}$. The XUV flux at the
planetary orbit is $F_{\rm XUV} = \frac{L_{\rm XUV}}{4\pi d_0^2}$,
where the stellar XUV luminosity depends on the bolometric
luminosity and the rotation period of the star (see
Figure~\ref{fig::mors_tracks}). $L_{\rm XUV}$ is commonly
considered as a sum of X-ray and EUV luminosities, where the
latter can be expressed through the former using empirical
relations \citep[see, e.g.,][]{sanz2011,Johnstone2020mors}. In
turn, $L_{\rm X}$ can be described as

\begin{equation}\label{eq::Lx_on_Prot}
    \frac{L_{\rm X}}{L_{bol}} = C(\frac{P_{\rm rot}}{\tau_{\rm
    conv}})^{\beta},
\end{equation}

\noindent where $\tau_{\rm conv}$ is the convective turnover time
depending on the stellar mass, and constants $C$ and $\beta$ are
different for the saturated and non-saturated regimes of the star
\citep[see, e.g.,][ and the Section 2.1 of the present
paper]{wright2011,Johnstone2020mors}. Thus, $F_{\rm XUV} \sim
\frac{L_{\rm bol}}{d_0^2}\times (\frac{P_{\rm rot}}{\tau_{\rm
conv}})^{\beta}$. We adopt the convective turnover time from
\citet{Spada2013} following the stellar models by
\citet{Johnstone2020mors} used in this work. This value varies
with time, but most of the changes happen before the age of
$\sim$100~Myr for the stellar mass range of 0.2-1.1~\Msun
\citep[see Appendix A of][for details]{Spada2013}. As the
dependence of the coefficients in Equation~\ref{eq::fat_tanh} on
$F_{\rm XUV}$ is relatively weak compared to, e.g., $T_{\rm eq}$,
it is possible to simplify this dependence and set $\tau_{\rm
conv}(age\geq 100{\rm Myr}) \simeq const$, and take as constant
the value of $\tau_{\rm conv}$ at the time of 500~Myr. For the
stellar masses considered here these values are $\tau_{\rm conv} =
$ 31.7, 60.4, and 135 days for 1.0, 0.75, and 0.5~\Msun stars,
respectively (which leads to $\tau M_{*}^2 \simeq $ const). We
therefore do not consider ages smaller than 100~Myr in our
approximation. It serves two goals: simplification of $\tau_{\rm
conv}(age)$ described above, and the exclusion of the saturation
period in the stellar evolution, which allows to consider only one
value of $\beta$ in Equation~\ref{eq::Lx_on_Prot} within our
approximation. Additional motivation to exclude the first 100~Myr
when fitting the coefficients in Equation~\ref{eq::fat_tanh} is
that the relation $f_{\rm at}(R_{\rm pl})$ at this time is
affected by the initial luminosity of the planet, which is to a
large extent model-dependent.

We therefore consider the following planetary/stellar parameters
when parameterizing the coefficients $C_1$, $C_2$, and $C_3$ in
Equation~\ref{eq::fat_tanh}: planetary mass $M_{\rm pl}$,
normalised bolometric flux at the planetary orbit $\frac{L_{\rm
bol} [L_{\odot}]}{d_0^2 [AU^2]}$, Rossby number of the star
$R_{\rm o} = \frac{P_{\rm rot}}{\tau_{\rm conv}}$, and the age of
the system. As the dependence on the stellar mass becomes pretty
weak after inclusion of $\tau_{\rm conv}$, we leave it outside of
consideration.

With the parameters described above, we fit the coefficients in
Equation~\ref{eq::fat_tanh} as follows

\begin{eqnarray}
  C_1 &=& (0.0464)_{-0.0131}^{+0.0093} \times (\frac{L_{\rm bol}}{d_0^2})^{-0.09} \times R_{\rm o}^{0.04} \times (\frac{M_{\rm pl}}{M_{\oplus}})^{0.07}\nonumber \\ & & \times (\frac{age}{100 Myr})^{0.03} \label{eq::C1} \\
  C_2 &=& (1.0296_{-0.1676}^{+0.1304})\times (\frac{M_{\rm pl}}{M_{\oplus}})^{0.095}\times R_{\rm o}^{0.01} \times (\frac{L_{\rm bol}}{d_0^2})^{0.01} \label{eq::C2} \\
  C_3 &=& (1.0053_{-0.1753}^{+0.3647})\times (\frac{M_{\rm pl}}{M_{\oplus}})^{-0.04}\times R_{\rm o}^{0.02} \times (\frac{L_{\rm bol}}{d_0^2})^{-0.02}\nonumber \\ & & \times (\frac{age}{100
  Myr})^{-0.017} \label{eq::C3}
\end{eqnarray}

{In Figure~\ref{fig::approx} (bottom panel), we compare the
atmospheric mass fraction corresponding to the specific planetary
and stellar parameters to the one predicted by
approximation~\ref{eq::fat_tanh}. The plot contains $\sim 5500$
points, covering the whole range of planetary and stellar
parameters at 3 time snapshots: 100~Myr, 1~Gyr, and 5~Gyr; each
black cross represents a snapshot for an individual planet. For
atmospheric mass fractions above 0.5\% of planetary mass, 99.7\%
of the $f_{\rm at}/f_{\rm at, approx}$ points, given by central
values of coefficients $C_1$, $C_2$, and $C_3$, lie in between 0.5
and 2, and 88\% of points lie in between 0.8 and 1.2. The spread
decreases towards larger atmospheres, heavier planets, and older
ages. Thus, for planets heavier than 10~\Mer and atmospheric mass
fractions above 5\%, $\sim$82\% of points lie between 0.9 and 1.1
(in comparison to 63\% for the whole range of applicability). For
$f_{\rm at}\lesssim 0.3\%$, the approximation does not work
anymore and the relation $f_{\rm at}/f_{\rm at, approx}$ shows
clear systematic error (see the footnote at the beginning of this
Section).}

By analyzing the coefficients $C_1$, $C_2$, and $C_3$ and
comparing it to the scheme in Figure~\ref{fig::approx} (top
panel), one can see that the atmospheric mass fraction for a given
radius is larger for larger coefficients, and thus for smaller
bolometric luminosities (hence cooler planet) and slower rotation
of the star (hence smaller XUV flux). For larger planetary mass
(increase in $C_2$), the same atmospheric mass fraction results in
slightly larger radii for a lighter planet, with this difference
decreasing with increasing atmospheric mass. Finally, after ruling
out the dependencies on the stellar bolometric flux and rotation,
the remaining dependence on the age of the system (as a proxy of
the cooling planetary core) is relatively weak and predicts a
slightly larger $f_{\rm at}$ for the same radii at the later age
(and a cooler core). At the younger ages, not considered here,
however, the exponent in this dependence increases by about an
order of magnitude.

%%%%%%%%%%%%%%%%%%%%%%%%%%%%%%%%%%%%
\section{Conclusions}\label{sec::discussion}
The evolution of the atmospheres of low and intermediate mass
planets is closely connected to the physical properties of their
host stars. In this paper, we have modeled the evolution of a wide
range of sub-Neptune-like planets orbiting low-mass stars
(0.5-1~$M_{\odot}$), that follow different evolution paths. Our
main conclusions can be summarised as follows.

\begin{itemize}
    \item For all stellar masses, the distribution of the parameters of evolved planets shows similar patterns. Planets with larger masses
or orbiting farther out from the star keep more of their
primordial atmospheres than planets with smaller masses or
orbiting closer-in. Less straightforward, planets that start their
evolution having compact envelopes ($\sim3-5\%$ of their total
mass) tend to keep a larger fraction of their primordial envelope
for a given planetary mass and orbital separation.
    \item Additionally, planets evolve differently depending on the rotation
rate of the host star (i.e., its activity level). This difference
is more accentuated for planets orbiting more massive stars, as
the difference in $L_{\rm XUV}$ emitted by the slow and fast
rotating stars increases with stellar mass (see
Figure~\ref{fig::mors_tracks}).
    \item A planet has larger chances to keep its primordial atmosphere in
the habitable zone (or in any regions with equal equilibrium
temperature) of a  more massive star, even though luminosities of
low mass stars are lower. This fact is particularly relevant for
habitability studies.
    \item We have tested the influence of the initial (post-formation)
planetary luminosities on the parameter distribution of the
evolved planets by modeling three young planets, K2-33 b, AU Mic
b, and Kepler-411 d. We came to the conclusion that the uncertain
initial luminosity does not in general significantly affect the
state of the evolved planet. It can, however, constrain the
possibility to predict the future state of a young planet. To put
a tight constraint on the future evolution of the young planet
orbiting close to an active star, one needs a good observational
constraint on the planetary mass.
    \item We anticipate that all three planets considered in Section~\ref{ssec:model-entropy} will likely remain in the category of sub-Neptune-like planets after Gyrs of evolution. For planets experiencing relatively low atmospheric mass loss we can constrain the radii at 5~Gyr, that is $\sim$2.9-3.6~$R_{\oplus}$ and $\sim$2.9-3.2~$R_{\oplus}$ in the case of AU~Mic~b and Kepler-411~d, respectively. For K2-33~b, experiencing extreme atmospheric mass losses, we can only put a loosen constraint of $\sim$2.4-6.7~$R_{\oplus}$ due to the weak observational constraint on planetary mass. The upper limit of the radius can further slightly increase if the planetary mass is higher than 40~$M_{\oplus}$.
    \item In close-in orbits, where planets are exposed to extreme
atmospheric mass losses, having relatively small initial
atmospheric mass fractions (in range of $\sim$5-15\%) enhances the
chances for the planet to keep some of its primordial atmosphere.
This conclusion should be further investigated by
self-consistently including planetary formation models at the
initial condition of the simulations. However, already at this
point, we can conclude that certain combinations of initial
atmospheric mass fraction and post-formation luminosity of a young
planet at close-in orbit can allow it to keep a part of its
primordial atmosphere while at other starting conditions it would
not be possible (which could explain the presence of planets in
so-called hot-Neptune desert). It is thus clear that the
simultaneous modeling of the thermal evolution of the atmosphere
and the atmospheric mass loss, as well as an accurate prescription
of the starting parameters, is of crucial importance for very
close-in planets.
    \item Finally, we constructed an approximation describing the relation
between the atmospheric mass fraction and the radius of a planet
and bounded its coefficients to some of the main physical
parameters of planets and host stars. The analytical approximation
shown in equations \ref{eq::fat_tanh}, \ref{eq::C1}, \ref{eq::C2},
and \ref{eq::C3} can thus be used to quickly estimate the
atmospheric mass fraction for a planet of a given mass, at a given
age, orbiting a low-mass star at a given orbital separation.
\end{itemize}

\section*{Acknowledgements}
This project has received funding from the European Research
Council (ERC) under the European Union's Horizon 2020 research and
innovation programme (grant agreement No 817540, ASTROFLOW).

\section*{Data Availability}
The data underlying this article will be shared on reasonable request to the corresponding author.



%\bibliographystyle{mnras}


%%%%%%%%%%%%%%%%%%%%%%%%%%%%%%%%%%%%%%%%%%%%%%%%%%

%%%%%%%%%%%%%%%%% APPENDICES %%%%%%%%%%%%%%%%%%%%%

\appendix
\section{The relation between planetary luminosity and
entropy}\label{apx::S-L}

In general, the equation of state (EOS) defines univocally the
parameters of the system (matter) as a function of density and
temperature \citep[see, e.g.,][which is the basic EOS used in
MESA]{saumon1995}. The common approach in prescribing EOS is the
minimization of the Helmholtz free energy of a system given as a
function $F(V, T, \{N_{\rm i}\})$ of volume, temperature, and
particle numbers of specific species making up the ``fluid''
environment. In this formulation, the pressure and the entropy can
be expressed as the first partial derivatives of free energy:
$P(\rho, T) = -\frac{\partial F}{\partial V}|_{T,\{N_{\rm i}\}}$,
$S(\rho, T) = \frac{\partial F}{\partial T}|_{V,\{N_{\rm i}\}}$.
This pair of parameters dictate the mechanical ($P$) and
thermodynamical ($S$) equilibrium of the system (in our case,
planetary atmosphere), and appears explicitly in the EOS. Other
parameters, such as specific heats, expansion coefficients,
adiabatic gradients, etc., are defined as the secondary
derivatives by specific variables.

In MESA,  to create the initial model the solid core of a planet
(or a star) is defined by setting up the lower boundary
conditions. The user has to set up the mass ($M_{\rm c}$)
underlying the minimum (core) radius in the simulation (thus
setting the central pressure $P_{\rm c}$, or, equivalently,
density $\rho_{\rm c}$), and the temperature of the core.  For
stars, the latter is a specific value (low enough to exclude the
hydrogen burning). For planets, however, this approach does not
work if the initial guess on parameters is not close enough to
real ones \citep{paxton2013}. Thus for planets, the specific
temperature is defined through the EOS from the pair of $P_{\rm
c}$ and specific entropy $S_{\rm c}$, which are then adjusted
through the iteration procedure to attain the desired $(M_{\rm
pl}, R_{\rm pl})$. In mass coordinate, the luminosity depends on
entropy and temperature as \citep{paxton2013,Ushomirsky1998}

\begin{equation}\label{eq::apx_L(m)}
    L(m) \simeq -\frac{dS}{dt}\int_0^m dm'T(m')
\end{equation}

\noindent and the surface luminosity is estimated at the optical
depth of 2/3 as  $L(M_{\rm pl})  = 4\pi R_{\rm
pl}^2\sigma_{SB}T_{\rm eff}^4$. Further details can be found in
\citet{paxton2013}.

\begin{figure}
  \includegraphics[width=\hsize]{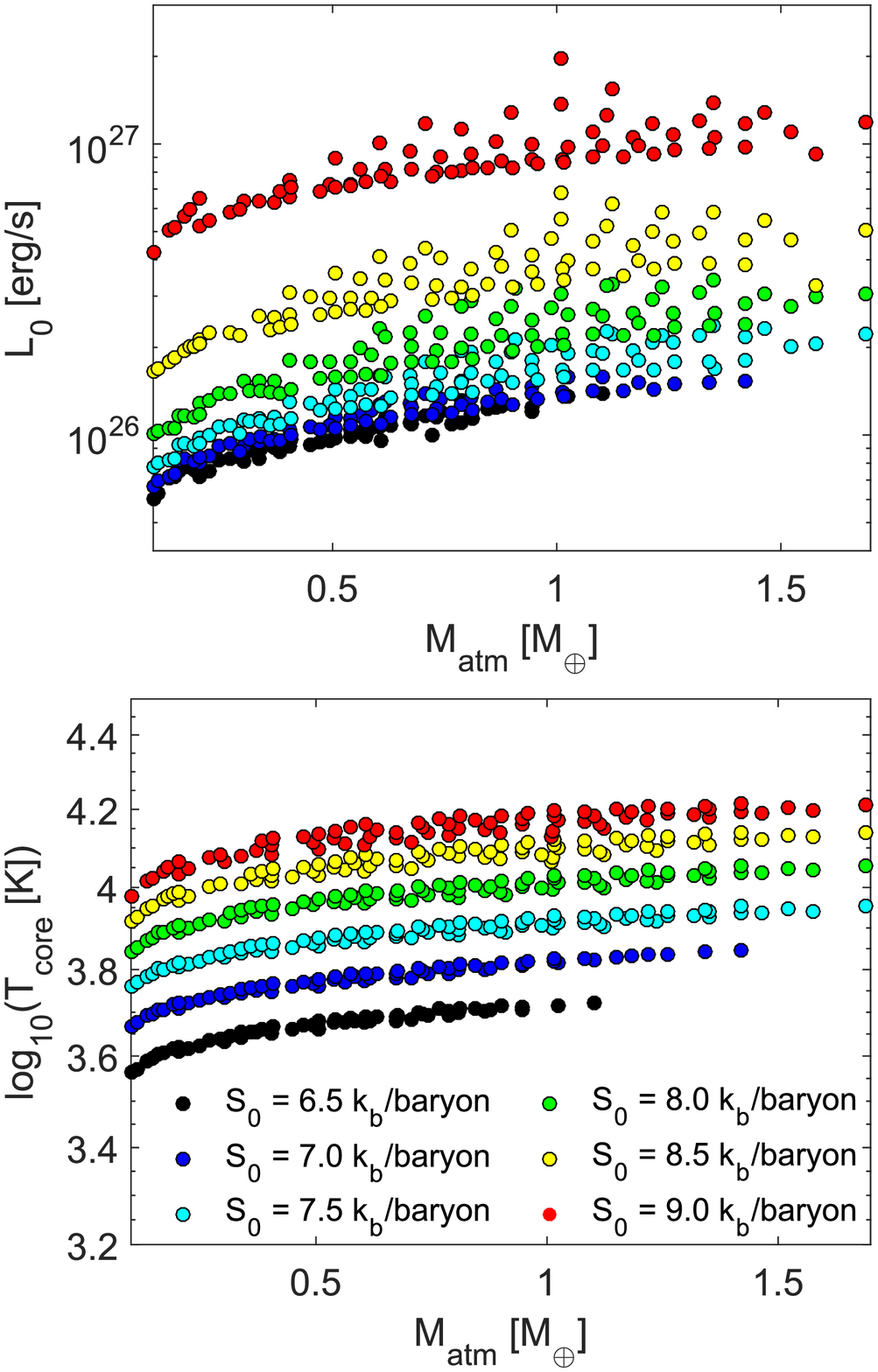}\\
  \caption{The initial planetary luminosity (top panel) and core temperature (bottom panel) of Kepler-411~d against the initial mass of the atmosphere as modelled in Section~\ref{ssec:model-entropy},
  for different levels of the initial entropy (color coded in the bottom panel).}\label{fig::apx_entropy}
\end{figure}

The specific entropy $S_{\rm c, 0}$, therefore, sets an
``amplitude'' for the temperature of the innermost cell in MESA
simulations (here referred to as $T_{\rm core}$) and the
luminosity of the planet. In Figure~\ref{fig::apx_entropy}, we
show the values of the initial luminosity and core temperature of
Kepler-411~d against the mass of the atmosphere for different
initial entropy levels. We only show here the case of Kepler 411
d, as the grid of the model points for this planet is the most
comprehensive.

%\section{Initial entropy distributions}\label{apx::S0}

\section{Population of evolved planets: atmospheric mass fractions}\label{apx::fat}

\begin{figure*}
  \includegraphics[width=\hsize]{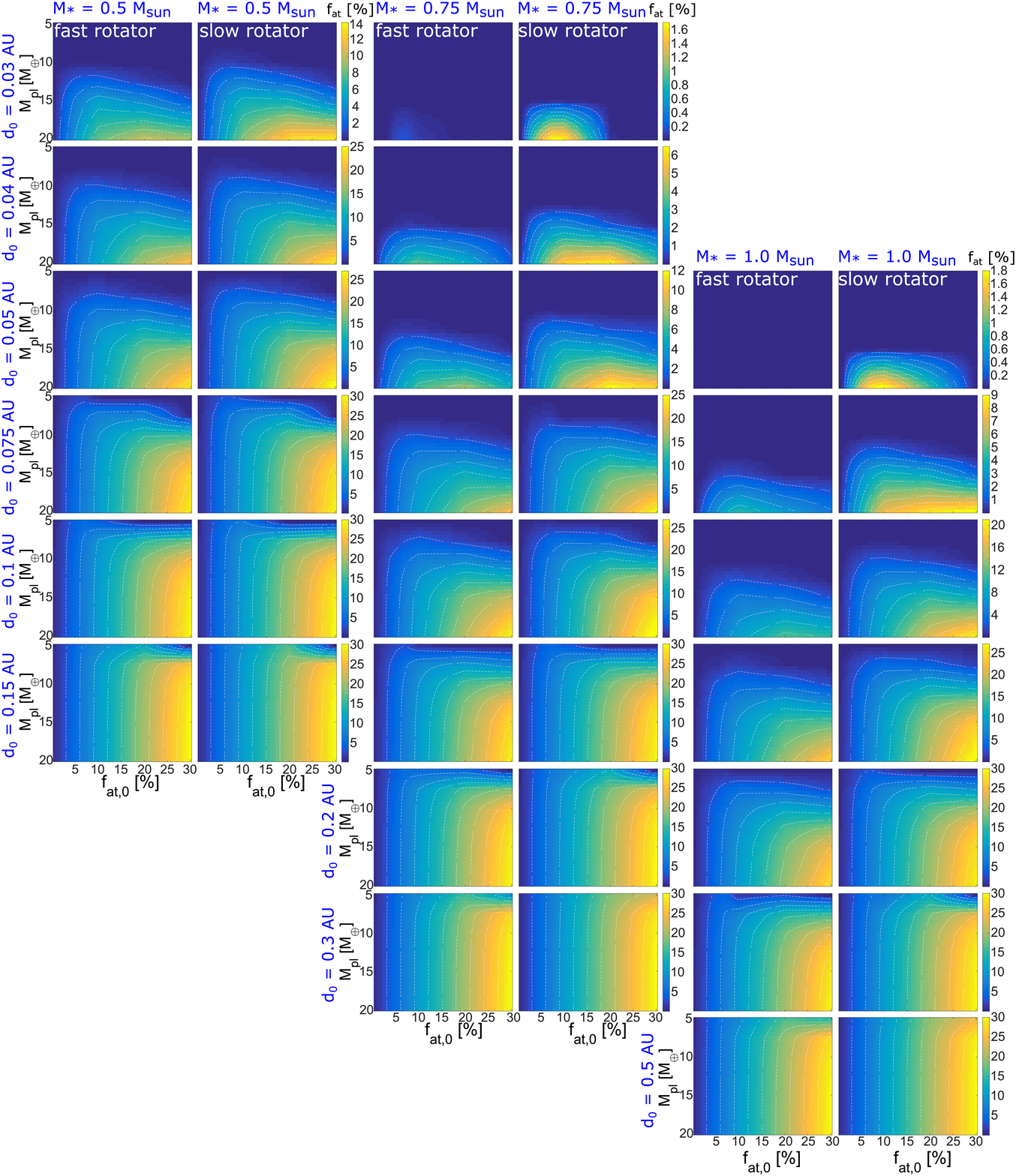}\\
  \caption{Atmospheric mass fraction at 5~Gyr of the test planets orbiting solar mass star evolving as the fast (left column) and the slow (right column) rotator.
  Orbital separations from top to bottom are 0.05, 0.075, 0.1, 0.15, 0.2, 0.3, and 0.5~AU. The 0.03 and 0.04~AU orbits are not shown, as all the test planets totally
  lose their envelops at these distances. }\label{fig::appx_fat}
\end{figure*}

Here we present the overall set of distribution of the atmospheric
mass fractions at the 5~Gyr age against the mass of the planet and
its initial atmospheric mass fraction, which has been outlined in
Section~\ref{ssec::results-overview}. The results are shown in
Figure~\ref{fig::appx_fat}, and organized after the host star
models: stellar masses of 0.5, 0.75, and 1.0~\Msun (pairs of
columns, from left to right), and rotation type (in each pair of
columns, the left one corresponds to the fast, and the right one
corresponds to the slow rotator). Different rows correspond to
different orbital separations between 0.03 and 0.5~AU (from top to
bottom). For lower mass stars (0.5 and 0.75~$M_{\odot}$), the
upper limit of $d_0$ is defined by the condition that the
equilibrium temperature should be above 300~K. For the solar mass
star, we omit the distributions at 0.03 and 0.04~AU, as all the
test planets lose their atmospheres entirely at these orbits, even
in case of the slow rotator.

\section{Population of evolved planets: radii of planets}\label{apx::rfin}

\begin{figure*}
  \includegraphics[width=\hsize]{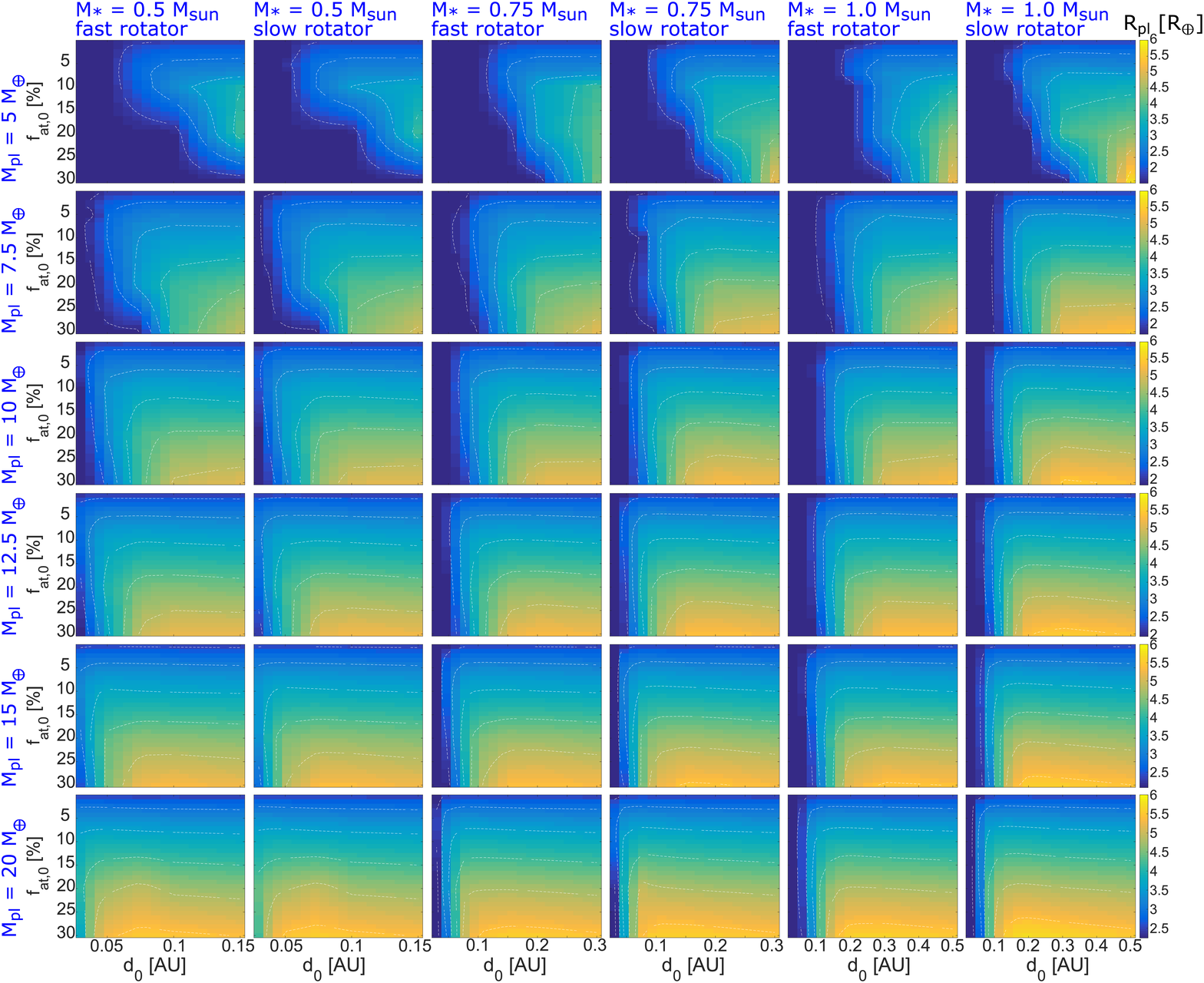}\\
  \caption{Radii of the model planets at 5~Gyr in dependence on the initial atmospheric mass fraction and orbital separation. Distributions are shown for different planetary masses from 5 to 20~\Mer (lines from top to bottom as
  referred in plots), and different host stars: 0.5~\Msun (two left columns), 0.75~\Msun (two middle columns), and 1.0~\Msun (two right columns). For each pair of columns, the left corresponds to the star evolving as the fast rotator,
  and the right column to the one evolving as the slow rotator. The radii of the solid cores assumed for the given planetary masses are, from top to bottom, 1.54, 1.72, 1.86, 1.98, 2.08, and 2.25~\Rer \citep{rogers2011core}.}\label{fig::appx_rpl}
\end{figure*}

To illustrate how the distribution of the atmospheric mass
fractions of evolved planets looks in terms of planetary radii, we
show in Figure~\ref{fig::appx_rpl} the distributions of planetary
radii against the initial atmospheric mass fractions (y-axis) and
orbital separations (x-axis). The columns are organised by stellar
models as in Figure~\ref{fig::appx_fat}, and the rows correspond
to different planetary masses from 5 to 20~\Mer (from top to
bottom). The minimum of the color scale (deep blue) for each
$M_{\rm pl}$ corresponds to the core radius of the simulated
planet. For low-mass planets, these distributions are mainly
shaped by escape, thus the effect of more compact atmospheres
surviving longer can be seen. The radius of a planet is mainly
defined by the atmospheric mass fraction and the temperature of
the planet. The core temperatures/luminosities of planets of
specific mass are nearly the same after 5~Gyr of evolution (see
the discussion in Section~\ref{ssec:model-entropy}). Thus, in the
parameter spaces where planets are weakly affected by the
atmospheric mass loss (i.e., at large $M_{\rm pl}$ or $d_0$) the
contour lines are nearly horizontal, with a small decay towards
larger orbital separations. The latter implies that the radii are
mainly affected by the internal heat of the planet, while the
external stellar heating has a minor effect.

%%%%%%%%%%%%%%%%%%%%%%%%%%%%%%%%%%%%%%%%%%%%%%%%%%

% Don't change these lines
\bsp    % typesetting comment
\label{lastpage}
\end{document}